\documentclass[twocolumn,secnumarabic,amssymb, nobibnotes,superscriptaddress, prb]{revtex4-1}
\usepackage{graphicx}
\usepackage{color}
\usepackage{multirow}
\usepackage{subfigure}
\begin{document}

\title{\emph{Ab initio}  study of the TiO$_2$ Rutile(110)/Fe interface}
\author{Anna Gr\"unebohm}%
\email{anna@thp.uni-due.de}
\author{Peter Entel}
\affiliation{Faculty of Physics and CENIDE, University of Duisburg-Essen, 47048 Duisburg, Germany}
\author{Heike C. Herper} 
\affiliation{Faculty of Physics and CENIDE, University of Duisburg-Essen, 47048 Duisburg, Germany}
\affiliation{Department of Physics and Astronomy, Uppsala University, Box 516, 751 20 Uppsala, Sweden}



%

\begin{abstract}
Adsorption of Fe on the rutile (110)-surface is investigated by means of {\it ab initio} density functional theory calculations. We discuss the deposition of single Fe atoms, an increasing Fe coverage, as well as the adsorption of small Fe clusters. It is shown that the different interface structures found in experiment can be understood in terms of the adsorption of the Fe atoms landing first on the rutile surface.
On the one hand, strong interface bonds form if single Fe atoms are deposited. On the other hand, the Fe-Fe bonds in deposited Fe clusters lead to a three-dimensional  growth mode. Mainly ionic Fe-oxide bonds are formed in both cases and the electronic band gap of the surface is reduced due to interface states.
Besides the structural and electronic properties, we discuss the influence of the interface on the magnetic properties finding stable Fe moments and  induced moments within the interface which leads to a large spin polarization of the Fe atoms at the rutile (110)/Fe interface.
 \end{abstract}
 \maketitle
\section{Introduction}
The (110)-surface of rutile  is one of the most important metal-oxide surfaces and has been subject of a large number of theoretical and experimental work in the last decades.\cite{Diebold, Zhang, Henderson,Fujishima}
Numerous applications ranging from thin film capacitors to optical devices are based on the exceptional dielectric properties of this oxidic-surface.\cite{Bordeaux,Fujishima}
The clean surface acts as a photocatalyst when exposed to UV-light. As the intensity of solar radiation on earth is maximal for visible light it is important to modify the frequency range of the photocatalytic response, which is possible by doping with transition metal atoms\cite{Teoh} and transition metal adatoms,\cite{Nolan} respectively. 
The use of adatoms is favorable as the 'spatial separation' of the excited holes and electrons improves the photocatalytic response.\cite{Nolan}
Adatoms modify the electronic structure, reduce the electronic gap and thus optimize the TiO$_2$ surface for the use in gas sensors \cite{Schierbaum} and solar cells.\cite{Graetzel}
In this context the Fe-rutile composite is important as it can be used for ammonia synthesis.\cite{Nobile}
In addition, the use of magnetic transition metal atoms such as Fe integrates the spin degree of freedom into the system, which allows for new functionalities.
For instance, the Fe/TiO$_2$ system has been discussed in the context of diluted magnetic semiconductors\cite{Co-DMS} and a magneto-electrical coupling has been detected in Fe-TiO$_2$ systems experimentally.\cite{Ishikawa}

General trends of low-temperature growth of (transition) metal atoms on the rutile surface are reviewed from an experimental point of view in Refs.~\onlinecite{Diebold,Diebold2} and references therein.
The growth mode of different metal atoms has been classified into two different modes.
On the one hand, group I and II metal atoms as well as transition metal atoms with a small number of valence electrons up to Cr interact strongly  with the TiO$_2$ surface.
Due to the large oxidability of these elements, charge from these adatoms is transferred to Ti at the surface. As a consequence, stable metal-surface bonds form which lead to rather flat metallic overlayers.
Also, the formation of ternary compounds is likely.
 On the other hand, if the oxidability of the metal atom is weak, e.g., for Cu or Zn, no charge is transferred to the surface and the metal-substrate interaction is weak.
 For this reason, no flat metal films  can be stabilized on the rutile surface.
 
The Fe-TiO$_2$ interaction is on the edge between these growth modes and the 
 energy of formation of Fe-TiO$_2$ composite systems is in the same range as the Fe-Fe interaction and the formation of Fe clusters.\cite{Hu}
Indeed, different growth modes of Fe on TiO$_2$ have been observed experimentally, depending on the preparation technique used  and growth parameters such as temperature and Fe flux.
The formation of flat Fe overlayers has been observed at low temperatures.\cite{Hu, Diebold, Pan}
In contrast, three-dimensional growth and Stranski-Krastanov growth (three-dimensional growth after the first layer has been formed) have been reported.\cite{Pan, Nakajima}\\
Therefore, a more profound understanding of the Fe-TiO$_2$ interaction, which can be obtained from {\it ab initio} simulations, is important in order to optimize the atomic and electronic interface structure. Although, the influences of temperature and growth kinetics are important,  simulations at $T=0$~K are capable to describe the basic mechanisms of the interface formation.
So far, systematic theoretical studies on the adsorption of nonmagnetic metals such as Mo,\cite{Asaduzzaman2} Pd,\cite{Murugan} Rh,\cite{Murugan} and single Cu, Ag and Au atoms exist.\cite{Giordano} 
However, with respect to iron, only the adsorption of single Fe atoms on two different surface positions ( which are energetically most favorable for adsorption of Mo and V) is discussed in Ref.~\onlinecite{Asaduzzaman}.
It has been predicted in that work that single Fe atoms can bind to atoms in different surface positions as long as three O neighbors (distance $<3$\AA) and an additional Ti bond towards the metallic adatom exist simultaneously.
The adsorption of Fe-O molecules has been discussed in Ref.~\onlinecite{Nolan}. It has been highlighted that different adsorption sites are thermodynamically stable and that the photocatalytic response is enhanced as the band gap is reduced by 0.3~eV because of Fe states near the valence band maximum.\\ 
In the present work, a more detailed study on the adsorption of single Fe atoms and a bundle of Fe atoms will be presented.
The latter setup is essential in order to model the growth mode at higher Fe flux or after the deposition of the first atoms.
We therefore present a first investigation of an increasing Fe coverage and of the adsorption of small Fe clusters on the TiO$_2$ surface in order to fill this gap.
It is shown that the Fe-Fe interaction reduces the surface wetting for an increasing number of Fe atoms deposited at the same time. In addition, the adsorption of the first Fe atoms is crucial for the further growth process and the interface symmetry. Thus, our results can be used to discuss many facets of experimental results. 
Furthermore, to our knowledge the magnetic properties of the rutile surface with adsorped Fe have not been discussed on a first-principles level so far. {We show that the magnetic moments of Fe are stable at the interface. In addition, finite magnetic moments are induced at Ti and O atoms in the interface via localized hybrid states. Our results suggest a similar magneto-electrical coupling as, e.g., discussed for Fe-BaTiO$_3$ interfaces in Ref.~\onlinecite{duan}. 

The paper is organized as follows.
In Section~\ref{sec:comp} we present the  computational methods used in this work which is followed by a discussion of the atomic and electronic structure of pure rutile and its (110)-surface in Section~\ref{sec:rutile}. 
The adsorption processes of single Fe atoms, which corresponds to the experimental setup of low-temperature adsorption at low Fe pressure, is discussed in  Section~\ref{sec:single}.
In Section~\ref{sec:mehr}, the influence of the Fe-Fe interaction on atomic and electronic structure of the Fe-TiO$_2$ interface will be discussed for larger Fe coverage and for attached clusters. The magnetic properties of the material are discussed in Section~\ref{sec:mag}.
Summary and conclusion can be found in Section~\ref{sec:conclusion}.
\section{Computational details}
\label{sec:comp}
The plane wave pseudopotential code VASP\cite{Kresse1} has been used for self-consistent electronic structure calculations using projector augmented wave potentials \cite{Blochl} and the generalized gradient approximation of Perdew, Burke, and Ernzerhof \cite{PBE}.  
It is a well known fact that the band gap of oxidic materials is underestimated by this approach. 
Partly, this failure can be attributed to the underestimation of the correlation between electrons in localized $d$-states, an error which can be corrected by the GGA$+U$ approach.
For oxygen deficient TiO$_2$, an onsite interaction of $U=5.5\pm0.5$~eV has been determined on an {\it{ab initio}} level.\cite{Calzado}
Choosing $U=5$~eV ($J=0$~eV) the influence on the atomic and electronic structure is tested by applying the GGA$+U$ approach in the formulation of Dudarev.\cite{Dudarev}
An energy cutoff of 500~eV and an energy convergency of $10^{-7}$\,eV guarantee highly accurate results. In addition, the relaxation of the ions was carried out until the forces were converged to 0.01~{\AA}/eV.
The $k$-mesh has been constructed with the Monkhorst-Pack scheme\cite{Monkhorst} and the number of points used  has been adjusted to the system size, e.g.,17$\times$17$\times$17 $k$-points have been used for bulk rutile whereas the mesh has been reduced to 4$\times$4$\times$1 for the investigation of Fe atom adsorption on the rutile surface.
A Gaussian-smearing of the electronic states of 0.1 and 0.05~eV has been used during the optimization of the atomic structure and for the determination of the electronic structure, respectively. 
  Local magnetic moments were obtained by projecting the wave functions onto spherical harmonics within spheres of 1.32~{\AA}, 0.82~{\AA}, and 1.30~{\AA} for Ti, O, and Fe atoms, respectively.
  
In order to reduce the computational effort, a minimal number of electrons has been treated as valence including the Ti $3d^4s^2$-, O $2s^2p^4$-, and Fe $4d^74s^1$-states. 
Although, the missing $p$ electrons in the Ti valence base may lead to small errors,\cite{Muscat} the main atomic relaxation as well as the most important features of the electronic structure and the polar character of TiO$_2$ are well accounted for by this approach.\cite{felich,Muscat,Bordeaux,Nolan}
 At least 10\,{\AA} of vacuum was used to prevent interactions between the periodically repeated slabs in case of free and Fe covered surfaces.
\section{Features of rutile}
\label{sec:rutile}
\begin{figure}
 \includegraphics[width=0.4\textwidth]{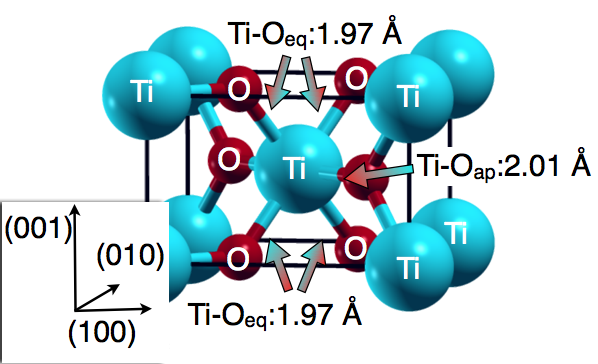}
  \caption{(Color online) (a) Structure of bulk rutile. Relative ion size corresponds to covalent radii. Light blue: Ti, dark red: O. Arrows mark the two different kind of Ti-O neighbors: shortest equatorial bond Ti-O$_{\text{eq}}$ and apical Ti-O$_{\text{ap}}$ bonds.}
  \label{fig:bulk}
\end{figure}
\begin{table*}
\caption{Lattice constants $a$, $c$, and internal parameter $u$, distances of the equatorial (TiO$_{\text{eq}}$) and apical (TiO$_{\text{ap}}$) Ti-O neighbors in {\AA} (cf.\ Fig.~\ref{fig:bulk}), and static charge per atom $Q$. We note that the static charge is not a well defined quantity and thus different experimental and theoretical approaches yield quantitative different results.}
\begin{tabular}{ccccccccc}
\hline
\hline
&$a$ (\AA)&$c$ (\AA)&$u$&Ti-O$_{\text {eq}}$ (\AA)&Ti-O$_{\text{ap}}$ (\AA)&Q$_{\text{Ti}}$  $|e|$&Q$_{\text{O}}$  $|e|$&E$_{\text{gap}}$ (eV)\\
PBE&4.664\footnotemark[1]&2.969\footnotemark[1]&0.305\footnotemark[1]&1.96\footnotemark[1]&2.01\footnotemark[1]&1.99\footnotemark[1]\footnotemark[2]&-1.00\footnotemark[1]\footnotemark[2]&1.63\footnotemark[1]\\
PBE\footnotemark[3] &4.686\footnotemark[1]&3.076\footnotemark[1]&0.304\footnotemark[1]&2.01\footnotemark[1]&2.02\footnotemark[1]&2.11\footnotemark[1]\footnotemark[2]&-1.06\footnotemark[1]\footnotemark[2]&2.29\footnotemark[1]\\
PBE0&4.591\cite{Labat}&2.955\cite{Labat}&0.304\cite{Labat}&1.94\cite{Labat}&1.99\cite{Labat}&2.36\cite{Labat}\footnotemark[4]&-1.18\cite{Labat}\footnotemark[4]&4.05\cite{Labat}\\
Experiment&4.587\cite{Samara}&2.954\cite{Samara}&0.305\cite{Samara}&1.94\cite{Samara}&1.98\cite{Samara}&~2.6\footnotemark[4]\cite{Traylor}&-1.3\footnotemark[4]\cite{Traylor}&3.0\cite{Pascual}\\
Hatree Fock&-&-&-&1.95\cite{Reinhardt}&1.98\cite{Reinhardt}&2.75\cite{Reinhardt}\footnotemark[4]&-1.38\cite{Reinhardt}\footnotemark[4]~~&-\\
\hline
\hline
\multicolumn{8}{l}{\footnotesize\footnotemark[1]This work; \footnotemark[2]Bader\cite{bader_imp}; \footnotemark[3]$U=5$~eV; \footnotemark[4]Mulliken; \footnotemark[5]Fitting of the phonon dispersion;}
\end{tabular}
\label{tab:bulk}
\end{table*}
The rutile morphology of TiO$_2$ possesses P4$_2$/mnm symmetry. 
Beside lattice constant and c/a-ratio of the tetragonal distortion, the internal parameter $u$ determines the Ti-O distances. Each Ti atom is octahedrally surrounded by O atoms with four nearest equatorial neighbors (Ti-O$_{\text{eq}}$) and two next-nearest apical neighbors (Ti-O$_{\text{ap}}$), see Fig.~\ref{fig:bulk}.
The material properties obtained with the computational tools used here are in sufficient agreement with previous experimental and theoretical work, see Tab.~\ref{tab:bulk}. 
The atomic volume is slightly overestimated because of GGA potentials and the well known underestimation of the electronic gap is present. The description of the gap can be improved by the introduction of a Hubbard term for the Ti $d$-states, but simultaneously the description of the structural properties worsens.\\
The Ti-O bonds in the pure oxide are mainly ionic  but the formal ionicity of Ti$^{4+} $O$^{2-}_2$ is considerably reduced, as Ti $d$- and O $p$-states hybridize, {\it cf.} Ref.~\onlinecite{felich}.
 The  density of states (DOS) obtained in Fig.~\ref{fig:konv} agrees qualitatively with previous experimental and theoretical results.\cite{Courths, Bredow}
The valence band maximum consists of O $p$-states, which are oriented within the (100)/(010) plane, while a significant occupation of Ti $d$-states with $e_g$-character appears in the rest of the valence band.
The conduction band is composed of Ti $d$-states which possess mainly t$_{2g}$-character at the band minimum.
If a Hubbard $U$-term is applied, the conduction band minimum is shifted to higher energies and the  hybrid peak which consists of O $s$-Ti $d$-states, 
is slightly shifted to lower energies.
However, no qualitative modification of the atomic and electronic structure has been observed if a $U$ correction is imposed. 

In the following, the properties of the TiO$_2$ (110)-surface, which are essential for the adsorption process, will be summarized. We refer to Refs.~\onlinecite{Bredow, Bates, felich, Reinhardt} for further details of the atomic and electronic structure of this specific surface.
The (110)-surface is the energetically most favorable low index surface as it is charge neutral and the coordination of the surface atoms is close to the bulk values. Within each surface unit, only one Ti and one O atom are undercoordinated as rows along the [001]-direction of 5-fold and 6-fold coordinated Ti atoms alternate (Ti$_{\text{5c}}$/Ti$_{\text{6c}}$), see Fig.~\ref{fig:slab}. 
In the remainder of the paper the nomenclature Ti$_{\text{5c}}$/Ti$_{\text{6c}}$ will be used for the Ti atoms in the surface layer 1 and the corresponding atoms in the layers below.
 Charged TiO$^+_{\text{p}}$ and bridging O$_{\text{b}}^{-}$ layers alternate in [110]-direction and each surface unit, i.e. each monolayer (ML) consists of two  O$_{\text{b}}$ and one TiO$_{\text{p}}$ layer, see Fig.~\ref{fig:slab}. 
The surface is slightly buckled, as  the undercoordinated Ti$_{5c}$ and bridging O (O$_{\text{b}})$ atoms relax inwards whereas Ti$_{\text{6c}}$ and O atoms within Ti-O planes (O$_{\text{p}}$) move outwards, see arrows in Fig.~\ref{fig:slab}.
The Ti-O distances along the surface normal alternate with the surface distance, whereas the modifications of the Ti-O distances within the surface plane are minor.\cite{Bredow}
The underlaying mechanism for these modifications of the Ti-O bonds along the surface normal
is the reduced coordination of O$_{\text{b}}$ and Ti$_{5c}$  atoms in the surface layer.
For these atoms, the energy difference between Ti $d$- and O $p$-orbitals, which are aligned along the surface normal, is reduced, the hybridization and the bond strength increase resulting in reduced Ti-O distances.\cite{Bredow, Schelling, felich} 
\begin{figure}
\centering
  \includegraphics[width=0.3\textwidth, clip, trim=4.2cm 5.2cm 10cm 0cm]{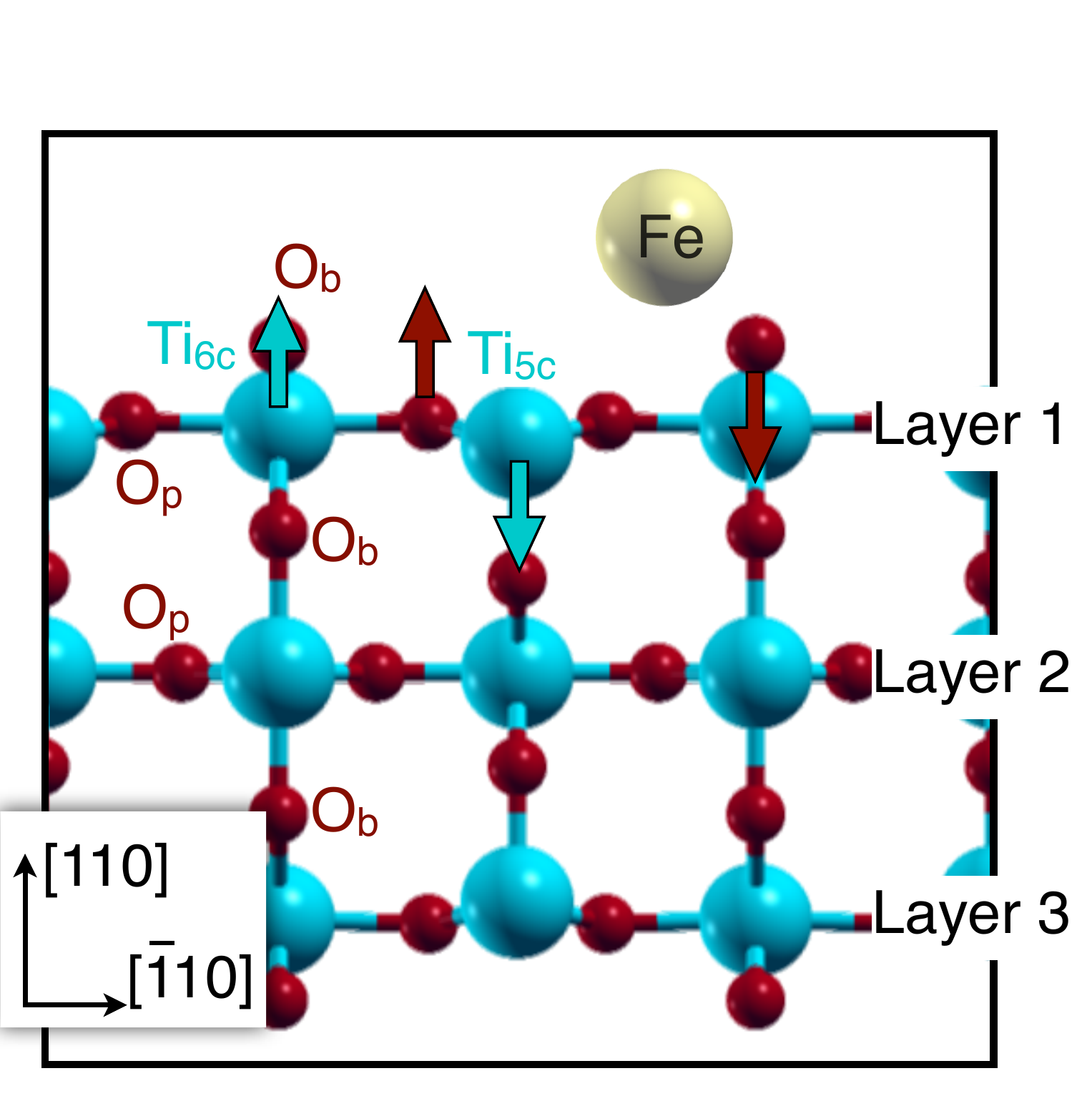}
  \caption{(Color online) Atomic structure of a free 3 monolayer thick rutile (110)-film. Arrows illustrate the atomic relaxation at the surface.  Relative ion size corresponds to covalent radii. Ti: light blue,  O: dark red. The most favorable adsorption position for a single Fe atom is indicated as well.}
  \label{fig:slab}
 \end{figure}
\begin{figure}
\includegraphics[width=0.5\textwidth,clip, trim=1.5cm 1.2cm 13cm 0.7cm]{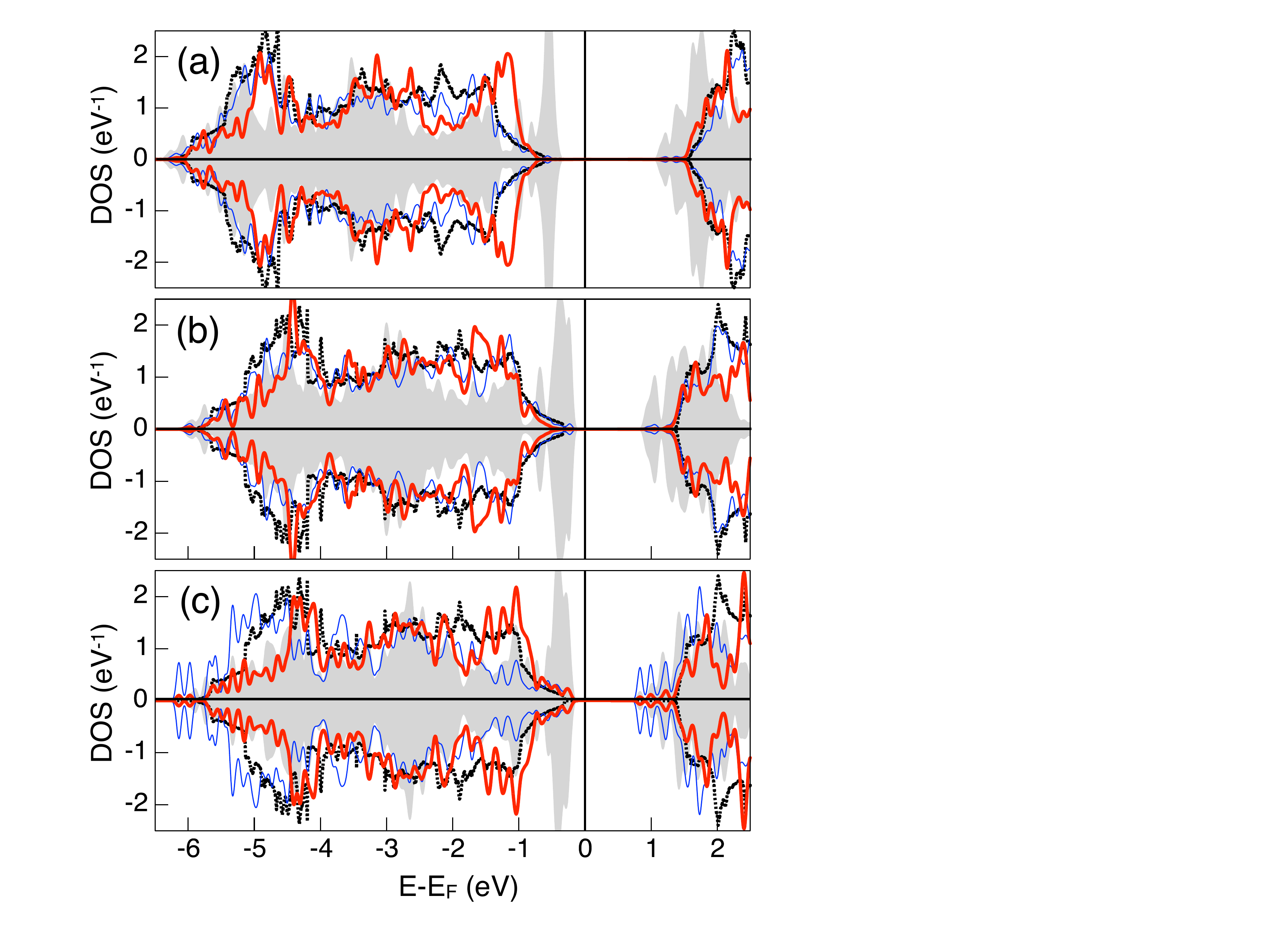}
\caption{(Color online) Layer resolved density of states (DOS) per TiO$_2$ formula unit. (a) 5 monolayers TiO$_2$ with two fixed bottom layers with $U=5$~eV; (b) as (a) with $U=0$~eV; (c) 3 monolayers with one fixed bottom layer  for the case $U=0$~eV. 
In each case the black dotted line corresponds to the bulk reference. Light gray: Bottom layer; Thin blue line: central layer; Thick red line: surface layer.}
  \label{fig:konv}
 \end{figure}
This modification of the bonds is accompanied by a charge transfer along the surface normal, see Tab.~\ref{tab:rhoob}.
For example, the charge of the topmost O$_{\text{b}}$ atom is reduced by 14\% (17\%) within our Bader analyis (the Mulliken analysis in Ref.~\onlinecite{Reinhardt}), while the charge of the Ti atoms in the first layer increases by 2\% (4\%).
This means that local dipole moments develop along the surface normal.
Since an overall dipole moment along the surface normal would destabilize the surface, a stability saving opposite charge transfer takes place for the Ti-O bonds below, which results in reduced Ti-O hybridization and enlarged Ti-O distances for the corresponding Ti-O bonds.
\begin{table}
\caption{Static charge distribution per atom [$|e|$] within the topmost free layers for different film thicknesses for free films and films with fixed bottom layers. In the last row the tight-binding charge distribution for a free film is listed for comparison}
\begin{tabular}{cccccc|ccccc}\hline
\hline
&\multicolumn{5}{c}{Layer 1 (surface)}&\multicolumn{5}{c}{Layer 2 (surface -1)}\\
ML&O$_{\text{b}}$&O$_{\text{p}}$&Ti$_{\text{6c}}$&Ti$_{\text{5c}}$&O$_{\text{b}}$&O$_{\text{b}}$&O$_{\text{p}}$&Ti$_{\text{6c}}$&Ti$_{\text{5c}}$&O$_{\text{b}}$\\
\hline
3\footnotemark[1]\footnotemark[4]&-0.86&-1.01&1.96&1.96&-1.03&-0.98&-1.00&1.98&1.99&-0.95\\
5\footnotemark[1]\footnotemark[4]&-0.86&-1.02&1.96&1.94&-1.00&-0.98&-1.00&1.98&1.98&-0.96\\
5\footnotemark[2]\footnotemark[4]&-0.93&-1.06&2.08&2.08&-1.10&-1.05&-1.06&2.11&2.11&-1.04\\
5\footnotemark[1]\footnotemark[5]&-0.82&-1.06&1.96&1.99&-1.01&-0.95&-0.99&1.92&1.99&-0.95\\
9\footnotemark[1]\footnotemark[5]&-0.86&-1.02&1.96&1.95&-1.01&-0.99&-1.00&1.97&1.98&-0.96\\
\hline
5\footnotemark[3]\footnotemark[5]&-1.10&-1.17&2.35&2.29&-1.19&-&2.34&2.37&-&-\\
\hline
\hline
\multicolumn{11}{l}{\footnotesize\footnotemark[1]Bulk: 1.99/-1.00; \footnotemark[2]$U=5$~eV: Bulk: 2.11/-1.06; }\\
\multicolumn{11}{l}{\footnotesize\footnotemark[3]Ref~\onlinecite{Schelling}; Bulk +2.36/-1.18; \footnotemark[4]Fixed bottom layers; \footnotemark[5]Fully relaxed.}
\end{tabular}
\label{tab:rhoob}
\end{table}
Due to the large polarizability of TiO$_2$, the atomic surface relaxations  discussed here extend several ML into the surface and are thus truncated for thin films, and hence induce finite-size effects in the electronic structure. 
These finite size effects open up the possibility of "fine-tuning" of the electronic structure of the TiO$_2$ surface by using ultrathin films.
In order to separate the general trends of the Fe-TiO$_2$ interface from the thickness dependent modulations, a detailed discussion of the influence of film thickness on the atomic and electronic structure is necessary.\\
The atomic relaxations of ultrathin TiO$_2$ films depend on whether the films possess a center of inversion (odd number of layers) or not (even number of layers). In the first case, the amplitude of the discussed oscillation of the interlayer distances shrinks with increasing film thickness, while it increases for an even number of layers.
In order to reduce the influence of the parity of the number of layers, it is convenient to fix the ionic positions at the bottom.
Here, we restrict ourselves to the discussion of an odd number of layers and present a systematic comparison of completely relaxed films and films with fixed bottom layers.\\
\begin{table}
\caption{Left: Deviations of the Ti-O$_{\text{eq}}$ distances compared to a 9 ML film with a fixed bottom layer (\%). Right: electronic band gap as function of on the number of layers $N$ without $U$ correction. 
}
\begin{tabular}{ccccc|cccc}
\hline
\hline
&\multicolumn{4}{c|}{Distances}&\multicolumn{4}{c}{E$_{\text{gap}}$}\\
&\multicolumn{2}{c}{Layer 1}&\multicolumn{2}{c|}{Layer 2}&\\
$N$&Ti$_{\text{6c}}$&Ti$_{\text{5c}}$&Ti$_{\text{6c}}$&Ti$_{\text{5c}}$&Total&Surface&Ref.~\onlinecite{Murugan}&Ref.~\onlinecite{Bredow}\footnotemark[1]\\
\hline
3~~&~0.5/-1.0&-0.5&0.8&-2.7/3.3&1.05&1.05&-&-\\
3\footnotemark[2]&~0.5/-1.0&-0.8&1.0&~-3.5/4.2&1.20&1.20&1.27&1.45\\
5~~&0.3/-0.2&-0.3&0.0&-0.5/0.7&1.12\footnotemark[3]&1.36\\
5\footnotemark[2]&~0.3/-0.2&-0.3&0.3&-1.2/1.2&1.43&1.43&1.58&1.71\\
7~~&~0.0/~0.0&~0.0&~0.0&~0.0/~0.3&1.19&1.64&-\\
7\footnotemark[2]&~0.0/~0.0&-0.3&~0.0&-0.4/~0.4&1.51&1.51&-&1.77\\
9\footnotemark[2]&~0.0/~0.0&~0.0&~0.0&-0.1/0.2&-&-&-&1.85\\
\hline
\hline
\multicolumn{9}{l}{\footnotesize\footnotemark[1]Bulk: E$_{\text{gap}}=1.93$~eV; \footnotemark[2]Fully relaxed; \footnotemark[3]With U: E$_{\text{gap}}=$1.54~eV.}\\
\end{tabular}
\label{tab:relax}
\end{table} 
The Ti-O$_{\text{b}}$ distances which we obtained for different film thicknesses are listed in Tab.~\ref{tab:relax}.
For both, free films and fixed bottom layers, all Ti-O$_{\text{p}}$ distances as well as the Ti-O$_{\text{b}}$ distances within the surface are sufficiently converged with respect to the thickness for 3 ML. Similarly,  the charge state of the topmost O$_{\text b}$ is nearly independent of the film thickness, see Table~\ref{tab:rhoob}.
For free films, the convergency of the atomic charges with respect to the film thickness is much slower. In this case more than 5 ML are required in order to obtain converged charge distributions. Although, the overall ionic character of the Ti-O bonds is enlarged by a $U$ term, the spatial distribution of the charge and its convergency with respect to the number of  layers is practically not modified when $U$ is taken into account.\\
As the topmost Ti-O bonds are most important for the adsorption processes,  qualitative results can already be expected for ultrathin film of 3 ML thickness.
However, the Ti-O distances along the surface normal in the center of the film show considerable deviations between films of 3 and 9 ML thickness.
As a consequence, the electrostatic potential of the central Ti$_{\text{5c}}$ atom is modified for the free film which results in a considerable reduction of the band gap within the whole film, as Ti$_{\text{5c}}$ states below the conduction band edge are induced,\cite{Murugan, Bredow} see Fig.~\ref{fig:konv}.
Although, the gap opens up with the number of layers, it is still underestimated for 7 ML.\cite{Bredow}\\
If the symmetry of the film is broken by fixing the bottom atoms, no such overall reduction of the bandgap appears. 
But, artificial surface states are induced for the fixed atoms. Most notably, a Ti$_{\text{5c}}$-state below the conduction band minimum appears in the fixed layer which has mainly $d_{x^2-y^2}$-character 
and hybridizes with the Ti$_{\text{5c}}$ atom above.
Furthermore, a Ti-O hybrid peak appears at the valence band maximum. The energetically highest occupied states within this peak are $p$-states of the fixed O$_{\text{b}}$ atom. 
The overall gap is therefore smaller than for free films. However, the surface states decay rapidly with the distance towards the fixed surface and the convergency of the electronic structure in the free surface with respect to the film thickness is considerably improved.
Only for ultrathin films of 3 ML thickness,  hybridization between these artificial surface states and the states of the central layer appears, see Fig.~\ref{fig:konv}~(c). Thus, small deviations in the Fe adsorption process between thick films and a film of 3 ML thickness are possible.
Qualitatively, the discussed modification of the electronic structure with the number of layers is not modified if a Hubbard term is imposed on the Ti $d$-states.

In summary, we could show that the atomic arrangement and the charge distribution within the uppermost free layer have practically converged for a film of 3 ML thickness which is therefore sufficient for the study of basic adsorption mechanisms. In addition, the convergence of atomic and electronic structures can be improved as the bottom layers are fixed.

\section{Adsorption of a sub-monolayer of iron}
\label{sec:single}
 \begin{figure}
 \centering
   \includegraphics[width=0.45\textwidth,clip, trim=13cm 6cm 1.5cm 8.2cm]{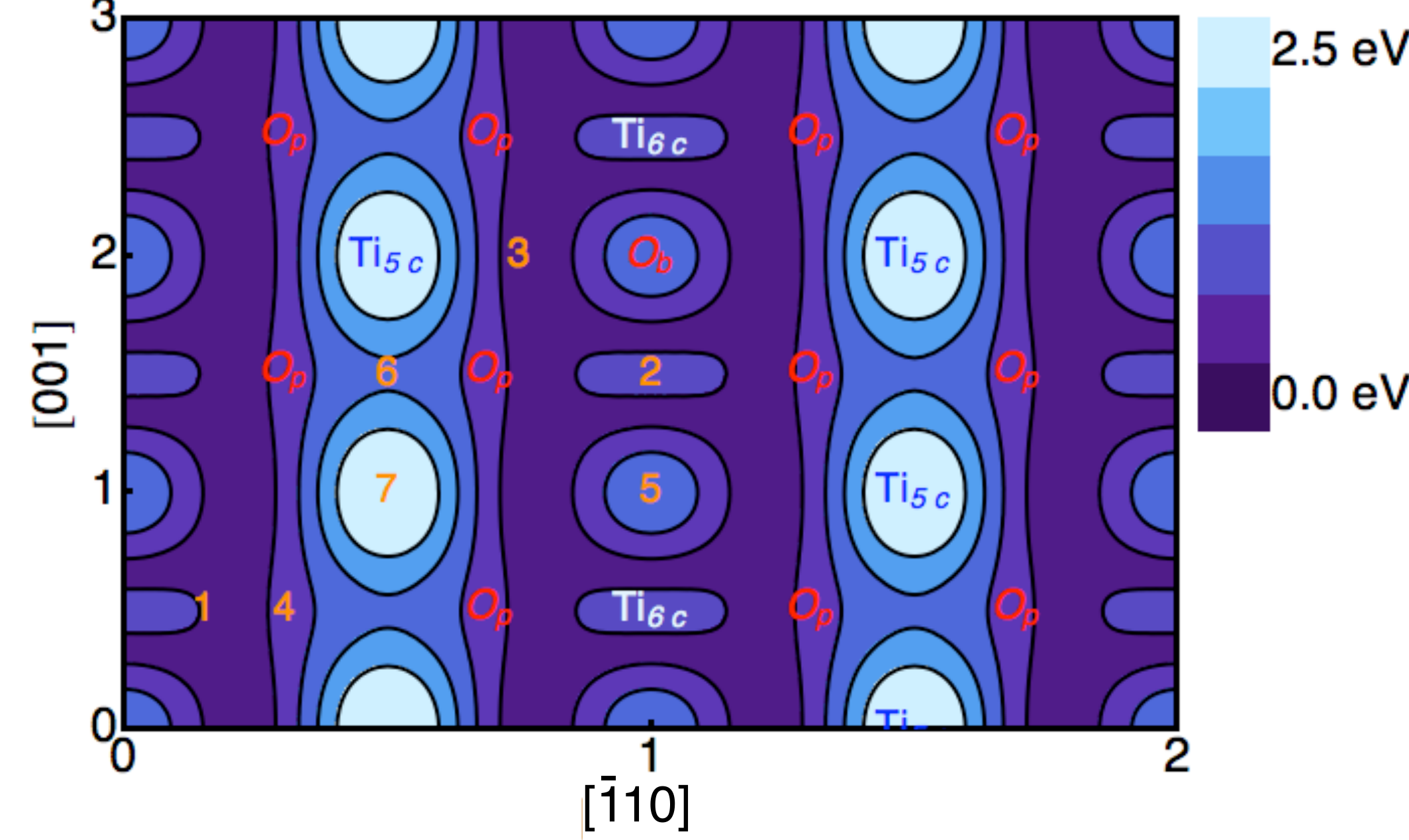}
  \caption{(Color online) Change in total energy for adsorption of Fe  on the rutile (110)-surface at positions 1-7  relative to the ground state (position 1), cf. Fig.~\ref{fig:slab}.}
  \label{fig:singleFe}
 \end{figure}
  \begin{table}
\caption{Characterization of adsorption positions $1$-$7$ in Fig.~\ref{fig:singleFe}. Listed are energy differences $\Delta E$ (eV/supercell) relative to position $1$, the change of atomic charges in the Fe neighborhood $\Delta Q$ ($|e|$), nearest neighbor Fe distances $d$ ({\AA}),  and magnetic moments $\mu$ ($\mu_{\text B}$/atom).}
 \addtolength{\tabcolsep}{-0.02cm}
\begin{tabular}{crccccccc}
\hline
\hline
&~~&1&2&3&4&5&6&7$^1$\\
\hline
\multirow{2}{*}{$\Delta E$}&&0.00&0.37&0.75&1.18&1.90&1.94&3.21\\
&&~\,0.00$^a$&-&~\,0.72$^a$&-&-&-&-\\
\hline
\multirow{7}{*}{$\Delta$ Q} &\multirow{2}{*}{Fe~~}&1.22&1.10&1.19&1.02&0.81&0.89&0.55\\
&&~\,1.16$^a$&-&-&-&-&-&-\\ 
&&~\,0.85$^b$&-&-&-&~\,0.71$^b$&-&~\,0.46$^b$\\
&Ti$_{\text{6c}}$&-0.08\,&-&-&-&-&-&-\\%
&Ti$_{\text{5c}}$&-&-&-0.05~\,&\,-0.07$^c$&-&\,-0.12$^c$$^d$&-0.10\,\\
&O$_{\text{b}}$&\,-0.14$^c$&\,-0.15$^c$&-0.12~\,&-0.14\,&0.13&-&-\\
&O$_{\text{p}}$&-0.14\,&-&-0.09$^c$\,&-0.05\,&-&\,-0.05$^c$&-\\
\hline
\multirow{6}{*}{$d$}&Ti$_{\text{6c}}$~~&3.20&2.72&\,2.91$^c$&2.67&~\,3.50$^c$&-&-\\
&Ti$_{\text{5c}}$~~&-&-&2.75&~2.8$^c$~~&-&~\,2.57$^c$&2.63\\
&\multirow{2}{*}{O$_{\text{b}}$~~}&\multirow{2}{*}{~2.50$^c$}&\multirow{2}{*}{~1.80$^c$}&\multirow{2}{*}{1.78}&\multirow{2}{*}{~2.07$^c$}&1.73&\multirow{2}{*}{~\,3.66$^e$}&\multirow{2}{*}{~\,3.5$^c$}\\
&&&&&&3.4$^c$\\
&\multirow{2}{*}{O$_{\text{p}}$~~}&\multirow{2}{*}{3.1~~}&\multirow{2}{*}{~3.2$^c$~~}&\multirow{2}{*}{~1.99$^c$}&1.93&\multirow{2}{*}{~\,3.9$^c$~~}&\multirow{2}{*}{~\,1.92$^c$}&\multirow{2}{*}{~\,3.0$^e$~~}\\
&&&&&2.86&&\\
\hline
\multirow{6}{*}{$\mu$}&Fe~~ &3.3~~&3.3~~&3.3~~&3.1~~&3.1~~&3.0~~&3.6~~\\
&\multirow{2}{*}{Ti$_{\text{6c}}$}&-0.1~~&-0.01&\multirow{2}{*}{0.01}&\multirow{2}{*}{-0.02\,}&\multirow{2}{*}{0.01}&\multirow{2}{*}{-}&\multirow{2}{*}{~0.01$^c$}\\
&&~0.5$^c$&0.4$^c$&&&&\\
&Ti$_{\text{5c}}$&-&-&-0.04~\,&-0.18\,&-$^f$&-0.13~&-0.06~\,\\
&O$_{\text{p}}$&0.08&-&0.06&0.04&-&0.06&~0.02$^e$\\
&O$_{\text{b}}$&~0.13$^c$&~0.15$^c$&-&0.06&0.08&-&-\\
\hline
\hline
\multicolumn{9}{l}{$^a$Ref.~\onlinecite{Asaduzzaman}; $^b$4~ML; $^c$two bonds of equal length;}\\
\multicolumn{9}{l}{\footnotesize $^d$0.05 electrons are transferred to Ti$_{\text{6c}}$ in the [001]-row;}\\
\multicolumn{9}{l}{\footnotesize  ($d$(Ti-Fe)$=$4.9~{\AA}); $^e$four bonds of equal length; }\\
\multicolumn{9}{l}{\footnotesize $^f$0.02$\mu_{\text{B}}$ are induced at all Ti$_{\text{5c}}$ atoms within the first layer. }
\end{tabular}
\label{tab:bader}
\end{table}
In this section we describe the modeling of the adsorption process of Fe atoms at low-temperatures and low Fe flux.
Under such experimental conditions individual Fe atoms are adsorbed on the surface and Fe-Fe interactions as well as diffusion of Fe atoms on or into the surface can be neglected. This process is modeled at zero temperature by posing single Fe atoms on the seven different adsorption positions of the rutile surface (cf. Figs.~\ref{fig:slab} and \ref{fig:singleFe}) and by relaxing the atomic positions under the constraint of a fixed bottom layer and fixed in-plane positions of Fe. In order to minimize the  interaction between periodic Fe images, a supercell of ($2\times3$) surface units is used which ensures a distance of 13.2~{\AA} (8.9~{\AA}) along [001] ([$\bar{1}$10]) between the periodic Fe images. The computational effort is minimized by using a minimal film thickness of 3 ML with fixed atomic positions in the  bottom layer.
In a similar manner the adsorption of single atoms on two different surface positions (which are energetically most favorable for Mo and V) has been discussed in Ref.~\onlinecite{Asaduzzaman}.
 The detailed energy landscape which has been obtained in the present investigation is shown in Fig.~\ref{fig:singleFe}.
An Fe atom attached between two O$_{\text{b}}$ surface atoms and one O$_{\text{p}}$ surface atom (position 1) is most favorable.
In this configuration, Fe has tetrahedral O environment which also occurs in various oxides.\\
The adsorption is 0.37~eV less favorable in case that Fe is placed above the Ti$_{6c}$ atom, with two O$_{\text{p}}$ neighbors (position 2).
For position 3 the Fe atom has two O$_{\text{p}}$ neighbors. This configuration is 0.75~eV higher in energy than position $1$. All other configurations are more than 1~eV higher in energy; whereby the most unfavorable position is on top of the Ti$_{5c}$ atom, see Tab.~\ref{tab:bader} and Fig.~\ref{fig:singleFe}.
Similar energy differences between position 1 and 3 have been obtained in Ref.~\onlinecite{Asaduzzaman} for a rutile film of 9~ML. This confirms the fast convergence of  adsorption energy with film thickness which has been predicted for the case of Pd adsorption, see Ref.~\onlinecite{Murugan}.\\
For all adsorption positions, the amount of charge ($\Delta$Q) is transferred from Fe to the substrate, in agreement with photoemission measurements of Diebold {\it {et al.}} where Fe$^{2+}$ has been detected on top of TiO$_2$. \cite{Diebold3}
This charge transfer is correlated with the adsorption energy, see Tab.~\ref{tab:bader}. 
The largest $\Delta$Q arises for position $1$, while $\Delta$Q is a factor of two smaller for the least favorable site $7$.
This correlation indicates the formation of ionic bonds between Fe and the surface, in agreement with the general trends of transition metal adsorption on TiO$_2$.\cite{Asaduzzaman}\\
However, no linear dependence between charge transfer and adsorption energy exists, e.g., $\Delta$Q of position 3 is similar to that of position 1 despite the energy  difference.
Most notably, different atomic relaxation energies due to the induced distortion of the TiO$_2$ structure disturb the linear dependence.
For example, only O$_{\text{b}}$ atoms next to Fe relax out of the surface for position 1 and thus the energy costs for the surface relaxation are minor. In contrast, a larger energy penalty for the surface relaxation has to be expected for site 3 as O$_{\text{b}}$ and O$_{\text{p}}$ atoms relax towards Fe. Similarly, the atomic rearrangements of position $5$ are restricted to the O$_{\text{b}}$ atoms next to Fe, whereas the Ti$_{\text{5c}}$ and O$_{\text{p}}$ 
\begin{figure}
\centering
\includegraphics[width=0.5\textwidth,clip, trim=1cm  5cm 8.5cm 1cm]{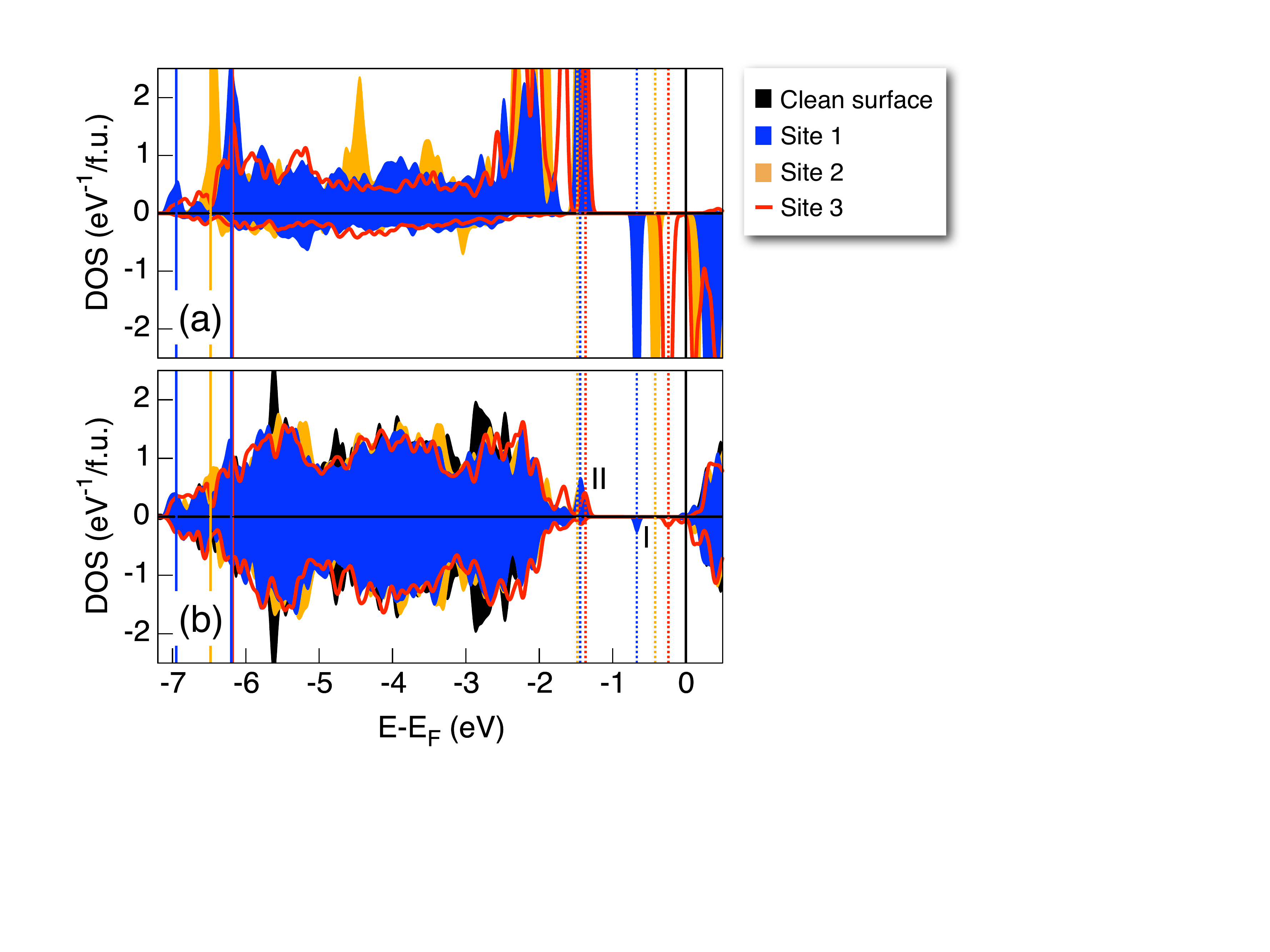}
\caption{(Color online) Modification of the density of states for different adsorption positions.
(a) Fe at different adsorption sites; (b) corresponding TiO$_2$ surfaces. Thin vertical lines mark Fe-Ti-O hybrid states for each configuration. The DOS of the free surface is shown as reference, which has been aligned with the lower valence band edge. \label{fig:singledos2}}%
 \end{figure}
atoms relax relative to each other for  position $6$.\\
 The adsorption mechanism can be understood by a closer look at the involved charge transfer and electronic states.
The charge state of O$_{\text{b}}$ calculated by Bader partitioning is reduced by 0.14 $|e|$ per atom at the clean rutile surface.
As a consequence, the electronegativity of these O$_{\text{b}}$ atoms is enhanced. This promotes the formation of ionic Fe-O$_{\text{b}}$ bonds accompanied by charge transfer to the film.
Adsorption of Fe atoms next to O$_{\text{b}}$ is therefore most favorable whereas position 7 is least favorable, since all Fe-O distances are larger than 3~{\AA} which prevents the formation of strong bonds. 
In the film, the injected charge spreads towards neighboring Ti atoms, see Tab.~\ref{tab:bader}.
In agreement, a reduction of the formal Ti$^{4+}$ state to Ti$^{3+}$  has been observed experimentally. \cite{Diebold3}\\
 Figure~\ref{fig:singledos2} illustrates the modification of the DOS for adsorption positions 1-3.
Within the entire valence band Ti-Fe and O-Fe hybrid states form. The states in the lower part of the valence band are marked by solid vertical lines.
For positions 1 and 3 Fe-O hybrid states form at the bottom of the valence band and  Ti-Fe$_{\text{6c}}$, Ti-Fe$_{\text{5c}}$ hybrid states at $\sim$-6.2~eV appear, respectively. 
Similarly, the next-nearest Fe$_{6c}$ atoms hybridize with Fe states resulting in a hybrid peak at  $-6.5$~eV for site 2. 
Besides, metal induced gap states (MIG) form in the gap of the rutile surface (states I below the conduction band  and states II above the valence band as marked by dotted vertical lines in Fig.~\ref{fig:singledos2})  in full agreement with experimental results, 
\begin{figure}
\centering
\includegraphics[width=0.5\textwidth,clip, trim=0cm 10cm 8cm 2cm]{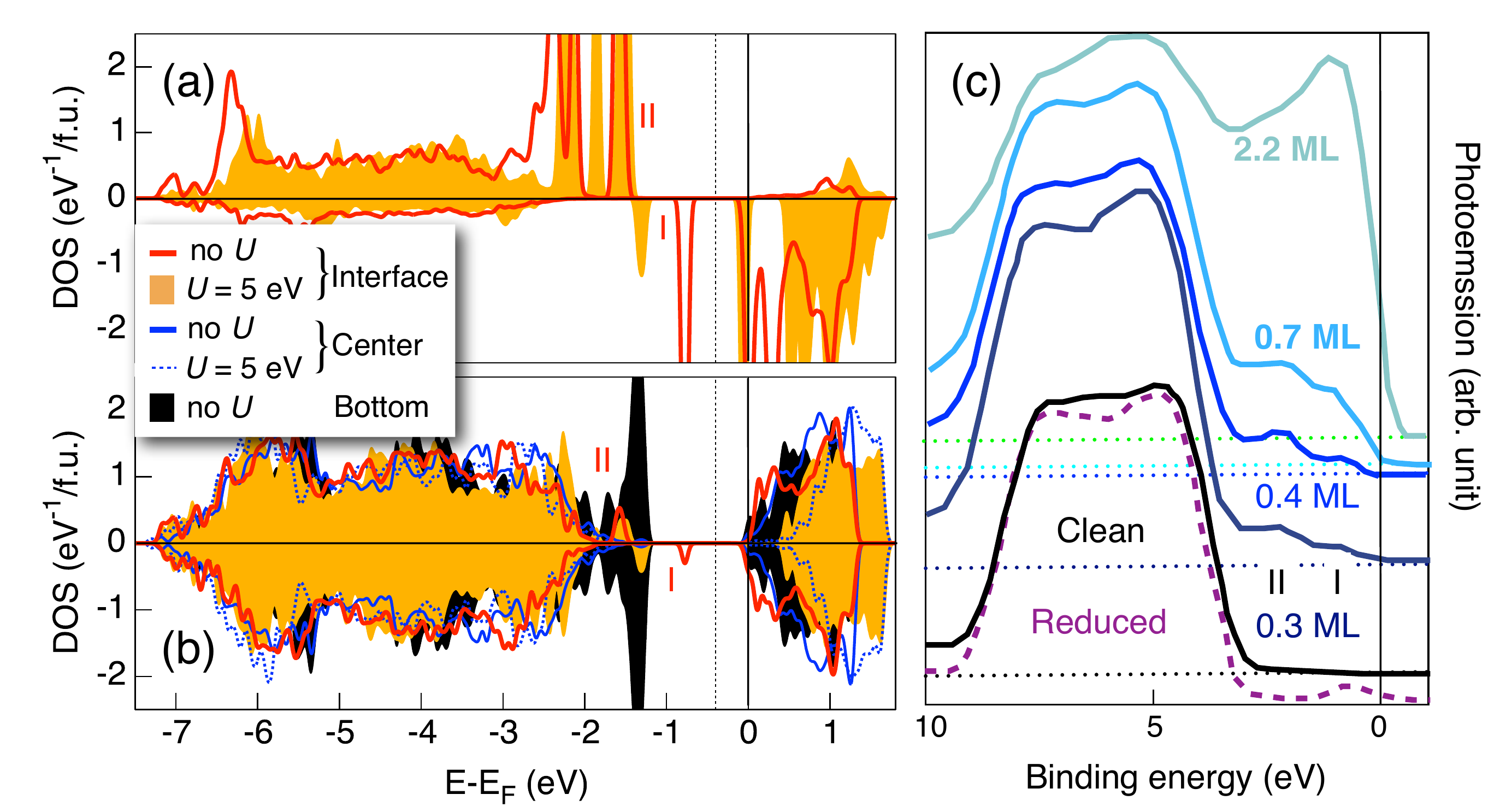}
\caption{(color online) (a)-(b) The influence of Hubbard correlations on the density of states for adsorption sites 1 (Fig.~\ref{fig:singleFe}) and a film thickness of 5~ML. (a) Fe and (b) TiO$_2$. Metal induced gap states (MIG)  I and II are marked.
The vertical dotted line marks the employed separation between free and bound charge, see text. \\
(c) Schematic sketch of the photoemisson spectrum adopted from Ref.~\onlinecite{Diebold3}. The spectrum of the clean surface (black) is opposed to a reduced surface (purple, dashed) and an increasing Fe coverage (shades of blue).
\label{fig:singledos}}
 \end{figure}
see Fig.~\ref{fig:singledos}~(c).\\
For sites 1 and 3 the Fe state I hybridizes with neighboring Ti$_{\text{6c}}$ and Ti$_{\text{5c}}$ atoms, respectively 
and a state is induced below the conduction band edge. This state can be assigned to a Ti$^{3+}$  surface state as obtained in experimental \cite{Diebold3} and theoretical  \cite{Calzado} investigations of the reduced rutile surface, see Fig.~\ref{fig:singledos} (c).\\
For Fe on adsorption site 2, the charge transfer towards Ti is much smaller  and no Ti defect state forms which could hybridize with the Fe state.
The differences in the Ti-Fe interaction and the corresponding charge transfer  can be traced back to the atomic arrangement. For sites 1 and 3 the neighboring O$_{\text{b}}$ and O$_{\text{p}}$ atoms mediate an indirect coupling between Ti and Fe and promote a large charge transfer.
In contrast, Fe is positioned directly above Ti$_{\text{6c}}$ at site 2, which enforces a direct Fe-Ti interaction without O mediated interaction. Such Fe-Ti interaction seems to be unfavorable and the charge transfer to the surface Ti atom is hindered. Analogously, no direct Fe-Ti$_{\text{5c}}$ interaction appears if the Fe atom is positioned above Ti$_{\text{5c}}$ (site 7).\\
A second MIG (state II) is visible at the valence band edge for all three adsorption sites, which is mainly of O $p$-type and localized at the O$_{\text{b}}$ neighbors,  see Fig.~\ref{fig:singledos2}. Thus the state observed experimentally in the same energy window in photoemission spectra\cite{Diebold3} is most likely an Fe-O hybrid peak. 
With increasing adsorption energy, the energy of this Fe-Ti$^{3+}$ state increases and approaches the valence band edge for the less favorable positions. Thus, the reduction of the band gap depends on the strength of the interface bonds.\\
In summary, the interpretation of experimental results in Ref.~\onlinecite{Diebold3} that Fe bond charges transfer via an O-mediated charge transfer to interface Ti atoms, could be confirmed.\\
For all adsorption sites, approximately 60\% of the injected Fe charge is transferred to its neighboring atoms. 
In addition, free carriers occupy the lowest Ti states throughout the TiO$_2$ film and the valence states are shifted to lower energies, see Fig.~\ref{fig:singledos}. This charge spillout is a common failure of standard density functional theory potentials as the underestimated band gap enforces a wrong alignment of the Fermi energy at the metal-insulator  interface.\cite{Stengel2} 
 Besides, no major deviations of the upper valence states are enforced in comparison to the free surface, see Fig.~\ref{fig:singledos2}. 
Particularly, the gap below these spurious free carriers is conserved and the free and bound charges are well separated, e.g., at $\sim-0.4$~eV for site 1, see Fig.~\ref{fig:singledos}.\footnote{For some configurations no such unique separation is possible, e.g., for  position 3, see Fig.~\ref{fig:singledos2}.}\\
 As the gap depends on the thickness of the TiO$_2$ film ({\it{cf}}.\ Tab.~\ref{tab:relax}) also the magnitude of the charge spill-out differs, compare Fig.~\ref{fig:singledos2} and Fig.~\ref{fig:singledos}. However, the wrong alignment cannot be prevented even if an even number of TiO$_2$ layers is used for which the gap is slightly larger than for bulk.\\
A more accurate modeling of the interface would require an improved approximation of the exchange correlation functional such as the hybrid functional B1-WC,\cite{Stengel2} which is out of the scope of the present work. 
The use of a Hubbard term allows for a qualitative cross-check of the obtained results without the spurious band alignment.
Indeed, the charge spill-out is reduced if $U=5$~eV is applied to the Ti $d$-states previous to the optimization of the atomic structure while the main features of the DOS are not modified, see Fig.~\ref{fig:singledos}. 
A more detailed discussion on the modification of the DOS is not  convenient as the value of $U$ has been determined for the deviating setup of clean reduced surfaces and is not necessarily transferable to the present setup.\cite{Calzado}\\
The hybrid states which have been discussed so far are localized at the interface and are not related to the spurious charge spill-out. Most notably, the description of the MIGs is in good agreement with experimental results, see Fig.~\ref{fig:singledos}. First of all, interface states I appear at $\sim-0.8$~eV and  $\sim-1$~eV below the Fermi level within the calculated DOS and the photoemission spectrum, respectively. Second, the interface states II are located directly above the valence band edge in both cases.\\
In summary, the wrong band alignment and the related charge spill-out may modify the details of the atomic and electronic structure at the interface. However, the main qualitative trends are  not affected.
\section{Increasing coverage of rutile (110) with iron atoms}
\label{sec:mehr}
\begin{figure}
\centering
\includegraphics[width=0.5\textwidth,clip, trim=2cm 5.5cm 4cm 2cm]{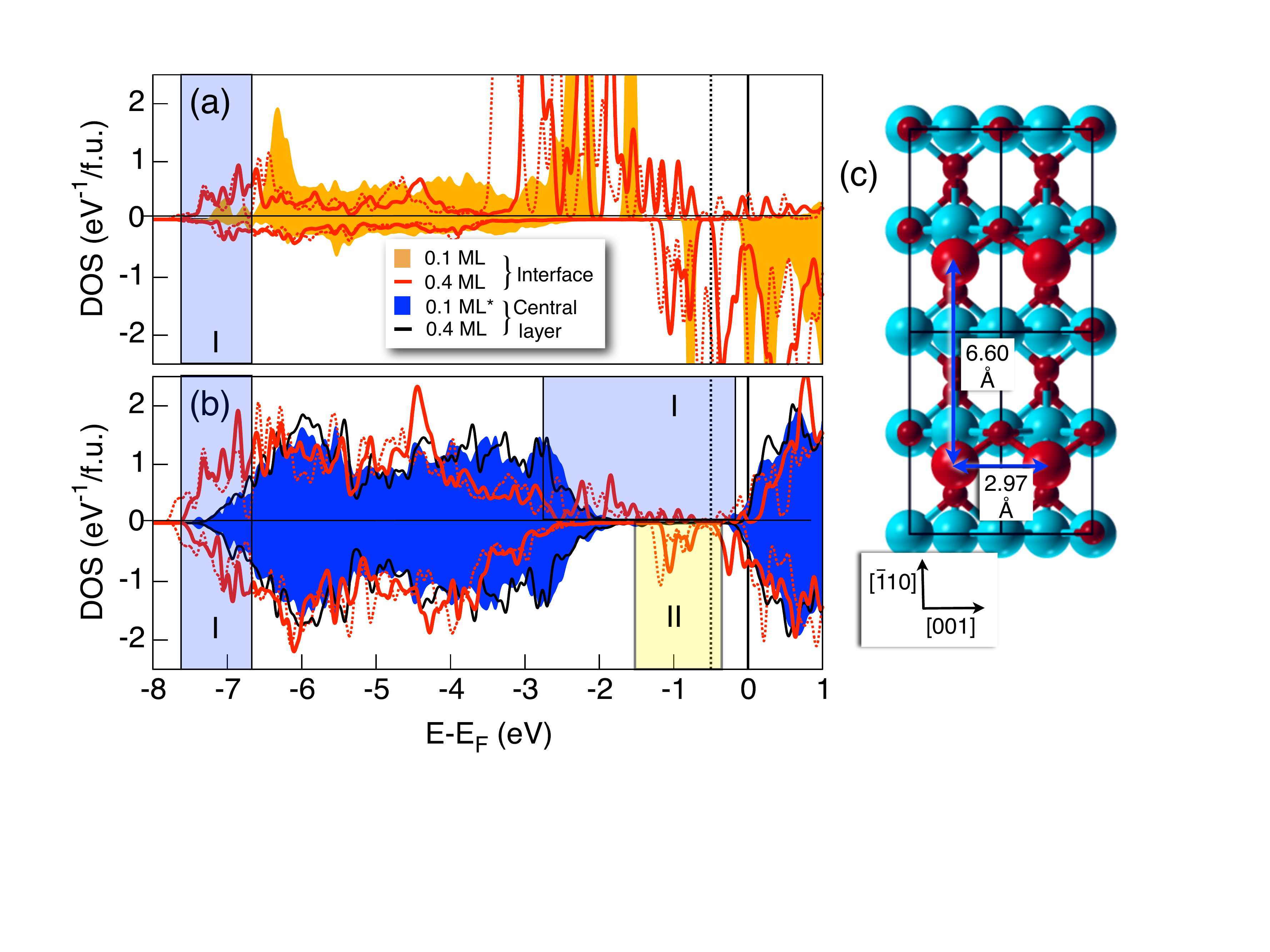}
\caption{(Color online)(a)-(b) Layer resolved density of states for  adsorption site 1 and different Fe coverages. A Hubbard term of 
 $U=5$~eV has been applied to the states marked by dotted lines. (a) Fe; (b) TiO$_2$; blue shaded region:  MIGs of type I and II (c) topview onto the interface. \\
 \footnotesize *States shifted by 0.2~eV to lower energies for better comparison. }
  \label{fig:densedos}
 \end{figure}
So far, the adsorption of single Fe atoms  has been modeled which corresponds to a coverage of  0.1~ML of bulk Fe. The Fe-Fe distances employed suppress Fe-Fe interactions. Besides charge spill-out and localized MIGs (Fig.~\ref{fig:singledos}), the insulating nature of the TiO$_2$ surface is conserved.
If the coverage increases to 0.4~ML, i.e., one Fe per ($1\times1$) surface unit, the Fe-Fe distances shrink to 2.9~{\AA} and 6.6~{\AA} along [001]  and [$\bar{1}$10], respectively.
 The atomic and electronic structure for this coverage is illustrated for site 1 in Fig.~\ref{fig:densedos}.
Since the nearest Fe-Fe distance is approaching its bulk value in this case, Fe states hybridize with each other.
The DOS peaks broaden and interface states I and II are spread over the main part of the gap, see the shaded region in Fig.~\ref{fig:densedos}.
As for the coverage of 0.1~ML, the states can be classified as O$_{\text{b}}$ and O$_{\text p}$-Fe hybrid states above the valence band edge (MIGs I) and Ti$_{\text{6c}}$-Fe hybrid states below the conduction band edge (MIGs II). All states in the shaded regions are localized in space and are qualitatively not modified by a Hubbard term.
Analogously, increasing weight and width of the interface states with increasing Fe coverage have been detected experimentally,\cite{Diebold3,Nakajima} see Fig.~\ref{fig:singledos}~(c).\\
The weight of the lower third of the valence band increases while states of the upper third are depopulated for increasing Fe coverage in agreement with experiment.\cite{Diebold3} For example, Fe-O$_{\text b}$ states 
in the shaded region around $-7$~eV gain weight and a Fe-Ti$_{\text{6c}}$ hybrid state builds up at $\sim-4.4$~eV, see Fig.~\ref{fig:densedos}~(a)-(b).  
The system gains energy by the formation or strengthening of hybrid states due to the adsorption of additional Fe and, hence, all states are shifted by $\sim0.3$~eV to lower energies.\\
By this mechanism, an increasing fraction of the lowest Ti$_{5c}$ states at the conduction band edge are occupied. Since the states in the central TiO$_2$ layer are not affected by this modified band alignment, see Fig.~\ref{fig:densedos} (b), no qualitative modifications of the interface properties due to the enhanced spill-out have to be expected.\\
The further increase of  coverage to  0.8~ML, i.e., two Fe atoms per surface unit, corresponds to a closed layer which is strained to the lattice constant of TiO$_2$.
\begin{figure}
\centering
\includegraphics[width=0.4\textwidth,clip, trim=0cm 16cm 18cm 2cm]{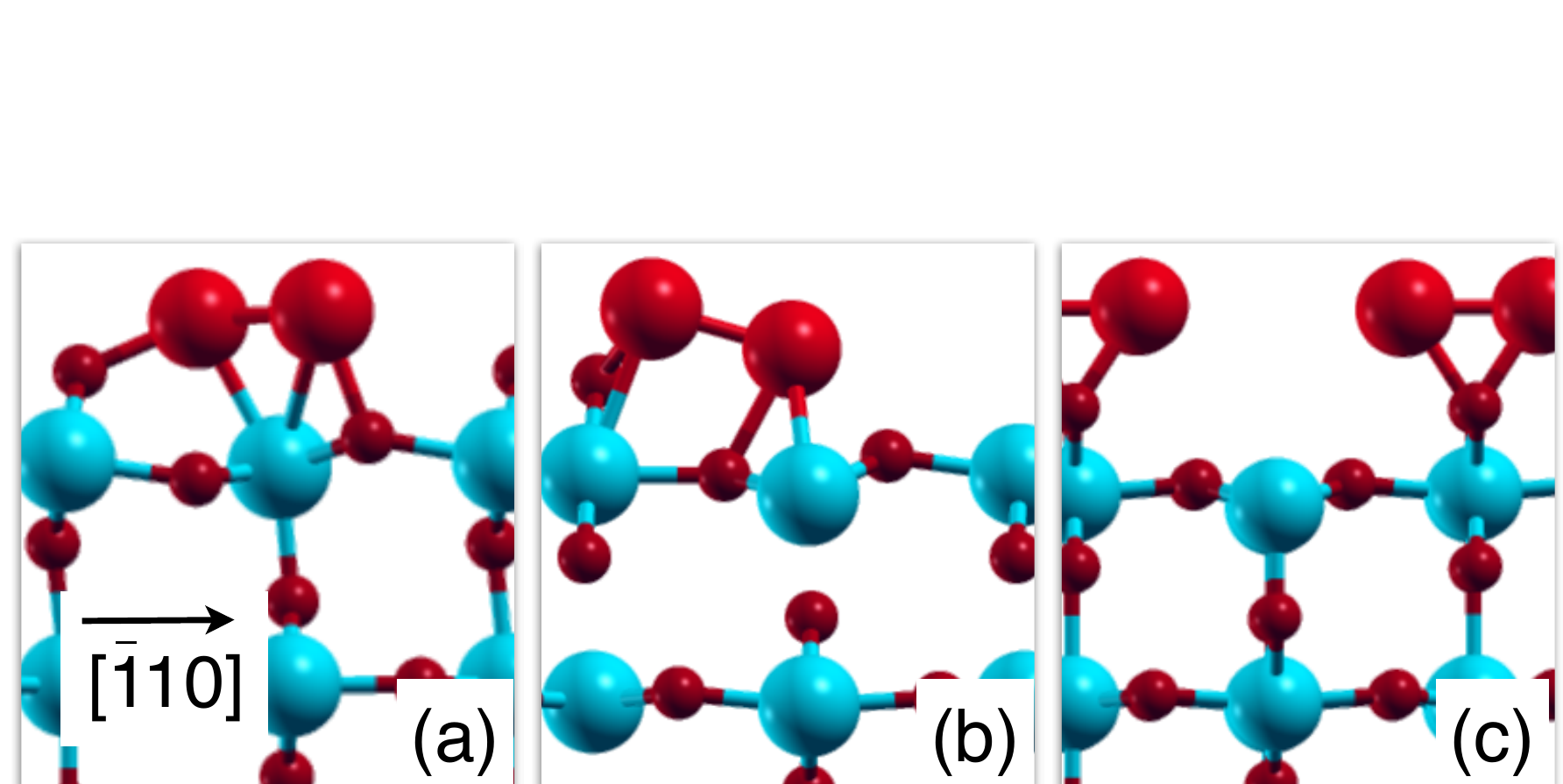}
\includegraphics[width=0.35\textwidth,clip, trim=18.5cm 13.5cm 0.2cm 2cm]{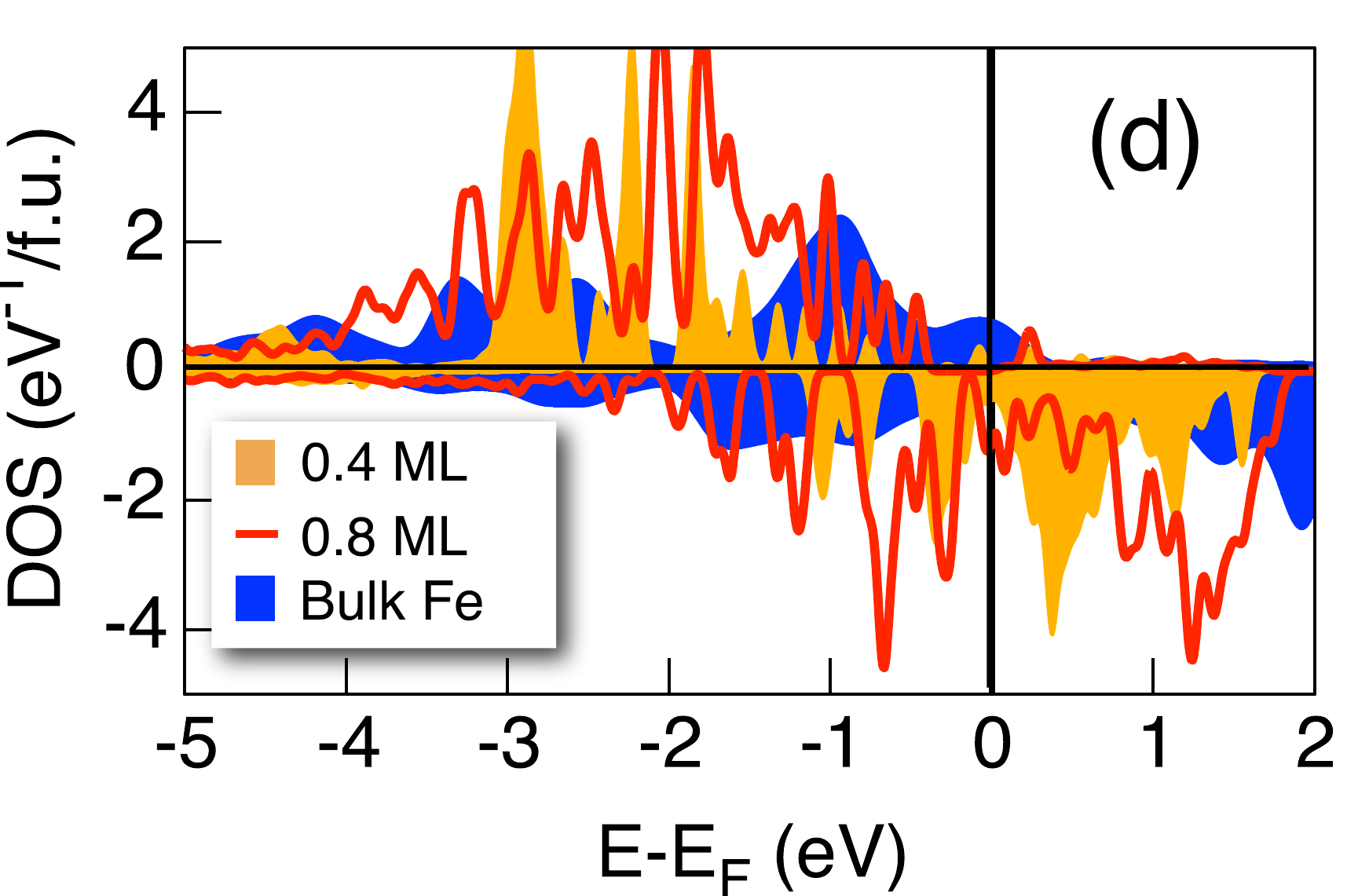}
\caption{(Color online)(a)-(c) Different adsorption scenarios for 0.8~ML Fe on rutile (110) (d) Density of states of Fe on rutile (110) for 0.4~ML Fe (position 1) and 0.8~ML Fe (configuration (a)) in comparison to bulk Fe.}
\label{fig:schicht}
\end{figure}
In order to investigate this coverage, the crystallographic orientation TiO$_2$(110)[1$\bar{1}$0]/Fe(001)[010]  is assumed which has been determined for Fe nanocrystals in a rutile matrix.\cite{Zhou}\\
 Three representative local minima of the energy landscape are shown in Fig.~\ref{fig:schicht}.\footnote{The Fe positions and the upper TiO$_2$ layers have been fully relaxed for different adsorption sites.} A homogeneous distribution of Fe is not favorable because of tensile strain of 16\% acting along [$\bar{1}$10]
  and because Fe atoms  reduce their [$\bar{1}$10]-distances.
This relaxation is most pronounced for configuration (c) for which both adatoms have initially been deposited next to site 3 (Fig.~\ref{fig:singleFe}). \\
After the formation of  Fe-O$_{\text b}$ and Fe-Fe bonds, the atoms are trapped in local minima of the energy landscape as all further Fe-O and Fe-Ti distances are larger than 3~{\AA}.
The Fe-Fe distance along [$\bar{1}$10] is the same range as for  the free dimer for which distances between 
\begin{table}
\addtolength{\tabcolsep}{-0.05cm}
\begin{center}
\caption[Interface distances of 0.8~ML Fe on rutile (110)]{Distances ({\AA}) of Fe towards its  nearest neighbors ($d<3$~{\AA}) for the  interface configurations in Fig.~\ref{fig:schicht} and relative energies in eV/(Fe atom).\label{tab:08}}
\begin{tabular}{ccccccc}
\hline
\hline
&$d$(Fe-Fe)$^a$&$d$(Fe-Ti$_{\text{5c}}$)&$d$(Fe-Ti$_{\text{6c}}$)&$d$(Fe-O$_{\text b}$)&$d$(Fe-O$_{\text p}$)&$E$\\
\multirow{2}{*}{(a)}&\multirow{2}{*}{2.37$^b$}&2.84$^b$ &\multirow{2}{*}{-}&-&2.00&\multirow{2}{*}{0}\\
&&2.54&&1.94&2.83$^b$&\\
\hline
\multirow{2}{*}{(b)}&\multirow{2}{*}{2.23}&2.68$^b$ &-&- &2.18$^b$&\multirow{2}{*}{0.40}\\
&&-&2.71&2.02$^b$&2.77&\\
\hline
\multirow{2}{*}{(c)}&\multirow{2}{*}{2.17}&\multirow{2}{*}{-}&\multirow{2}{*}{-}&\multirow{2}{*}{1.99}&\multirow{2}{*}{-}&\multirow{2}{*}{0.54}\\
&&&&&&\\
\hline
\hline
\multicolumn{7}{l}{\footnotesize$^a$Fe-Fe distances  along [001] are not listed; ($d=2.97$~{\AA});}\\
\multicolumn{7}{l}{\footnotesize$^b$Two interatomic bonds of equal length.}\\
\end{tabular}
\end{center}
\end{table}
2.01 and 2.26~{\AA}  depending on the spin configuration are given in literature.\cite{dissgeorg}\\
For configuration (b) the Fe atoms have initially been put at positions 2 and 6. An increasing amount of Fe-O and Fe-Ti bonds can form  which  lowers the total energy, although energy penalties have to be expected for the increasing Fe-Fe distance and the enhanced relaxation of the TiO$_2$ surface in comparison to (a).\\
If the Fe atoms are initially placed far from O$_{\text b}$ atoms the adsorbed atoms  are not trapped at the surface during the first relaxation steps and can thus arrange in a more favorable relative alignment, i.e., one additional Fe bond per surface unit is formed and each Fe atom possesses four Fe neighbors in comparison to 3 Fe-Fe  bonds for configuration (b) and (c), which results in a large energy gain of 0.5~eV.
The Fe positions correspond approximately to positions 3 and 6 after relaxation. In comparison to the adsorption of single Fe atoms, the Fe-O bonds are weaker due to the new Fe-Fe interaction i.e., the Fe-O distances increase, compare Tab.~\ref{tab:bader} and \ref{tab:08}.\\
Based on experimental results the formation of metallic Fe has been predicted  for a coverage of 0.7~ML\cite{Diebold3} and 3.0~ML,\cite{Nakajima} respectively.
 Although, Fe states start to form bulk like bands for a coverage of 0.8~ML, the pronounced Fermi edge of metallic Fe has not yet formed, see Fig.~\ref{fig:schicht}~(d).\\
Further increase of coverage is modeled for a layer-wise growth as was predicted in \onlinecite{Nakajima}.
For this purpose, 3 and 5 strained layers are deposited above the surface which corresponds to 2.4 and 4~ML of bulk Fe, respectively. The interface configuration shown in Fig.~\ref{fig:schicht}~(c) has been used as starting point for both coverages and the structures obtained by relaxation are shown in  Figure~\ref{fig:layers2}~(a)-(b). 
\begin{figure}
\centering
\includegraphics[width=0.5\textwidth,clip, trim=2cm 8cm 11.5cm 1cm]{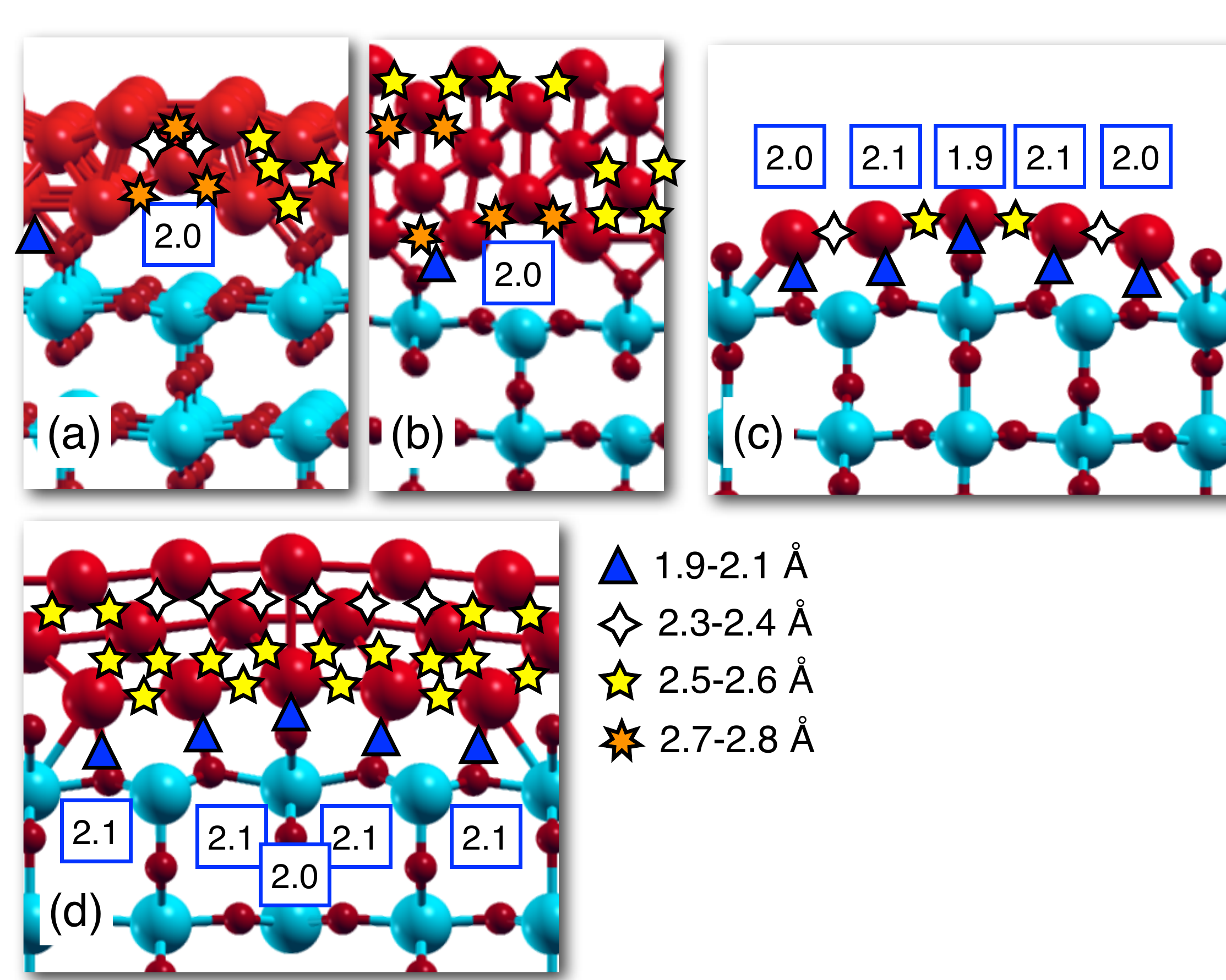}
\caption{(Color online) Atomic relaxation pattern for (a) 2.4~ML; (b) 4~ML; (c) 1~ML; (d) 3~ML Fe on rutile (110). Representative Fe-O and Fe-Fe distances are included. (a)-(b) Fe  is strained to the lattice constant of rutile; (c)-(d) Fe superstructure, see text. }
\label{fig:layers2}
\end{figure}
 Most notably, only two  O$_{\text b}$-Fe bonds per surface unit build up which possess a similar bond length as for 0.8~ML  coverage.
 This means that an energy barrier exists for further interface bonds, if the first Fe layer is adsorbed in the metastable state which is shown in Fig.~\ref{fig:schicht}~(c) and which cannot be overcome optimizing the structure at $T=0$~K.\\
 The tensile strain for three adsorbed layers in combination with the reduced coordination of the surface atoms leads to an inwards relaxation of the uppermost Fe layer and two buckled Fe layers form instead, compare  Figure~\ref{fig:layers2}(a).
By this relaxation, each Fe atom obtains either three or four nearest Fe neighbors with a distance of $d\sim2.3$-2.6~{\AA}, i.e., in the range of the bulk distance (2.45~{\AA}).
In addition, up to four next-nearest Fe-Fe bonds with $d\sim2.7$-2.8~{\AA} (bulk value 2.83~{\AA}) form. 
If the film thickness increases, the relaxation of the Fe layers is less pronounced as the ratio between surface atoms and bulk-like coordinated atoms in the center of the film decreases and a rather flat  film is stabilized, see  Figure~\ref{fig:layers2}(b).\\
Besides the formation of strained films, one may think of films which are not congruent with the underlying lattice constant along [$\bar{1}$10]. 
For example, a superstructure with a ratio of 2:5 between the TiO$_2$ and Fe lattice constant would reduce the epitaxial strain and the mean Fe volume is compressed by only 1.3\%.
\footnote{Due to the different symmetries of rutile and Fe inequivalent adsorption positions are occupied during the formation of a closed film anyway. }
The corresponding relaxed interfaces for 1 and 3~ML coverage are shown in Fig.~\ref{fig:layers2}(c-d). In both cases, the adatoms have initially been placed above Ti$_{6c}$ and  next to O$_{\text p}$  atoms in order to circumvent the unfavorable trapping near O$_{\text b}$ atoms. During relaxation, the Fe-Fe distances shrink and at least one Fe-O bond is formed for each Fe with bond lengths in the range of 2~{\AA} for a coverage of 1~ML. The Fe-O interaction is reduced, i.e. the Fe-O distances increase if the Fe-Fe coordination is increased by a coverage of 3~ML. Nevertheless, flat films with good surface wetting are obtained.\\
In summary, for an increasing coverage the interface structure  depends on the adsorption of the first deposited atoms which shows the influence of different kinds of growth parameters on the interfaces in experiments.
The formation of flat films with bcc-morphology is possible either by formation of strained films or by different lattice periods along [$\bar{1}$10] in rutile and Fe. 
\subsection{Adsorption of Fe clusters on rutile (110)}
\label{sec:cluster}
\begin{figure}
\centering
\includegraphics[width=0.5\textwidth,clip,trim=4cm 15cm 6cm 4cm]{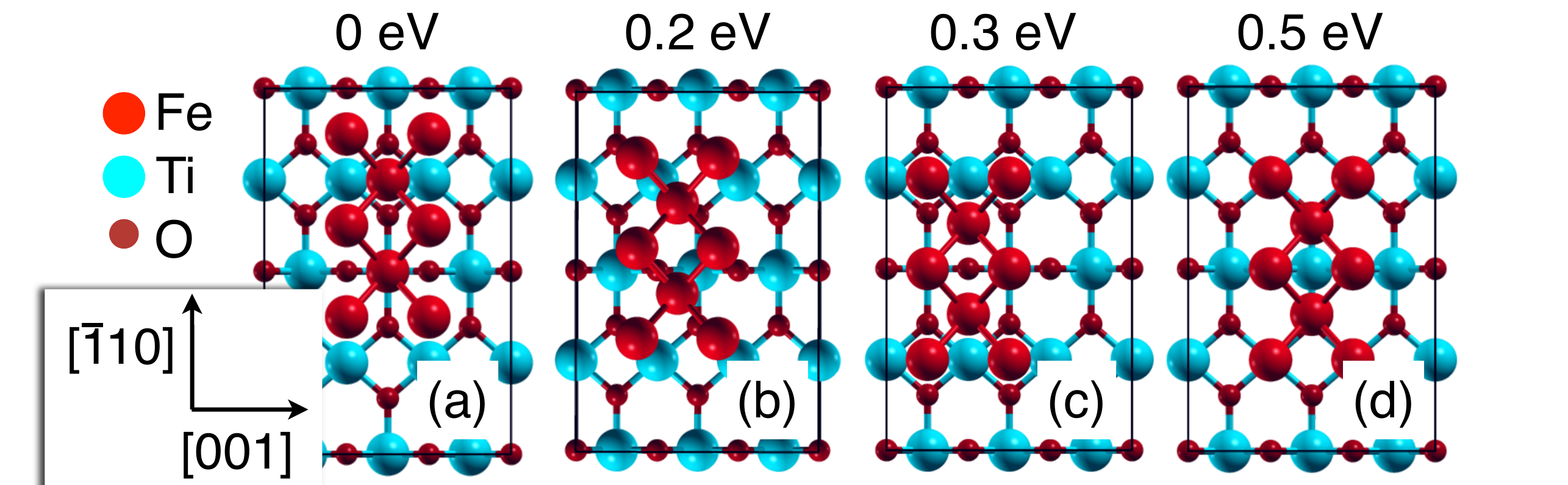}
\caption{(Color online) Topview on the initial configurations of a cuboid Fe$_{14}$ cluster attached to a  ($2\times3$) surface cell of rutile (110). Atoms within the direct surface are shown. The energy differences of the optimized structures relative to (a) are given in eV/(Fe atom).}
\label{fig:aufsicht}
\end{figure}
Complementary to the adsorption of single atoms, the case of clusters attached to the rutile surface is modeled in the following.
For this purpose, the adsorption of Fe$_{13}$ and Fe$_{14}$ clusters on the clean TiO$_2$ surface has been investigated.
In both cases, ($2\times3$) surface units are used, in order to reduce the interaction between neighboring clusters as schematically shown in Fig.~\ref{fig:aufsicht}.\\
The ground state of free Fe$_{13}$ clusters is a Jahn-Teller distorted icosahedron with FM alignment of all magnetic moments.\cite{prlgeorg}
If such a relaxed cluster is placed above the TiO$_2$ surface with its vertex pointing  towards the rutile surface, two adsorption trends can be distinguished.
If the vertex is initially positioned above the O$_{\text p}$-Ti$_{\text{5c}}$ region of the surface, the cluster aligns itself in between two O$_{\text{b}}$ rows and thus two strong Fe-O$_{\text b}$ bonds form at each side of the cluster ($d\sim1.9$~{\AA}) as shown Fig.~\ref{fig:cluster}(a).
For the second setup, the vertex of the cluster is initially placed near O$_{\text b}$ atoms, see Fig.~\ref{fig:cluster}(b). In this case, two Fe-O$_{\text{b}}$ bonds with a length of $d\sim 1.9$~{\AA} build up. In this case the system is trapped in a local energy minimum as has been discussed for the adsorption of films and a smaller number of interface bonds forms, e.g., two (one) Fe$_{\text{vertex}}$-Ti$_{5c}$ ($d\sim2.4$~{\AA})
and Fe$_{\text{vertex}}$-O$_{\text{p}}$ ($d\sim2.0$~{\AA}) bonds form for (a) and (b) in Fig.~\ref{fig:cluster}, respectively. Therefore, configuration (b) is  0.3~eV/(Fe atom) less favorable. \\
For both configurations, the relaxation of the TiO$_2$ surface is minor and restricted to direct neighbors of 
\begin{figure}
\includegraphics[width=0.5\textwidth,clip,trim=1cm 1cm 12.5cm 0cm]{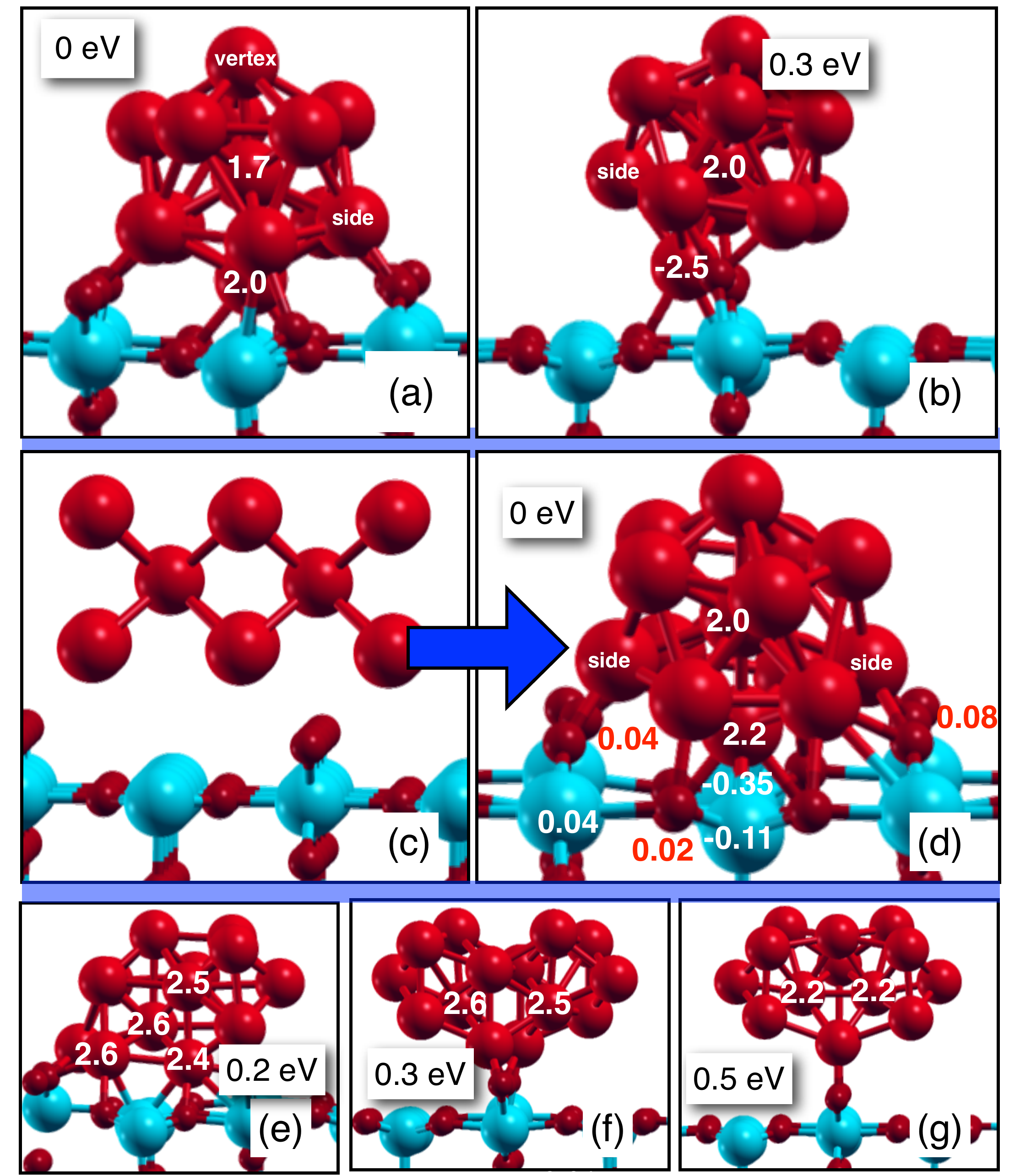}
\caption{(Color online) Adsorption of Fe clusters on rutile (110). (a)-(b) Icosahedral Fe$_{13}$
 (c)-(g) Cuboid Fe$_{14}$;
(c) Initial configuration corresponding to (d); (d)-(g) relaxed configurations sorted by relative energies (eV/ (Fe atom));  magnetic moments ($\mu_{\text{B}}$/atom) are given for Fe if the value differs from 2.8-3.0~$\mu_{\text{B}}$ and for representative Ti and O atoms.}
\label{fig:cluster}
\end{figure}
 Fe. Only the relaxation of one O$_{\text p}$ and O$_{\text b}$ atom towards the Fe cluster for (a) and (b), respectively, is significant.
For cluster (b) the whole cluster tilts towards the surface allowing for one Fe$_{\text{side}}$-O$_{\text{p}}$ ($d\sim2.0$~{\AA}), and two Fe$_{\text{side}}$-O$_{\text{b}}$ bonds ($d\sim2.0$ and 1.96~{\AA}).\footnote{The relaxation has only been performed up to 0.001 eV/step.}
Analogously, one Fe$_{\text{side}}$-Ti$_{5c}$ bond ($d\sim1.9$~{\AA}), and one Fe$_{\text{side}}$-O$_{\text p}$ bond ($d\sim1.9$~{\AA}) form in case of configuration (a).\\
In both cases, the icosahedral morphology of the cluster is conserved without  a tendency for surface wetting.
As the closed-shell  Fe$_{13}$ icosahedron is particularly stable, a large energy barrier for the fracturing of such cluster has to be expected, which cannot be overcome in the simulations at zero temperature.\\
In order to test whether this energy barrier or the intrinsic strength of Fe-Fe and Fe-TiO$_2$ bonds hampers surface wetting, the adsorption of  cubic Fe$_{14}$ clusters with bcc morphology has been investigated. Free clusters of this geometry are not favorable because of  the reduced coordination of Fe atoms in comparison with an icosahedral geometry and therefore the energy barrier for the redistribution of Fe may be overcome in $T=0$~K simulations.\\
As in the previous section, the crystallographic orientation TiO$_2$(110)[1$\bar{1}$0]/Fe(001)[010]  has been assumed. The cluster is strained to adopt the lattice constant of rutile, 
 and several adsorption positions have been sampled as shown in Fig.~\ref{fig:aufsicht}. The deposisted cluster is indeed not stable against rearrangement of Fe atoms, see Fig.~\ref{fig:cluster}(d-g). 
For all tried adsorption positions, the atomic relaxation shows similar behavior. First, Fe-O bonds form with a subsequent collapse of the clusters which results  in different  distorted clusters with local icosahedron geometry. For configurations (f) and (g), the cluster is again trapped in the local energy minimum next to the O$_{\text b}$ atoms. Furthermore, the simulation cells for these configurations possess mirror symmetry, which is conserved during the relaxation. This leads to a minor Fe-substrate interaction.\\
Although, the used conjugate gradient method does not reproduce the time-evolution of the system, the unique relaxation trends underline that Fe-O interaction is the driving force for Fe adsorption.\\ In particular, the low wetting tendency for an unfavorable cluster geometry shows that the formation of dense clusters is energetically more favorable than the  formation of further Fe-Ti or Fe-O bonds. Thus, a three-dimensional growth 
\begin{figure}
  \includegraphics[width=0.5\textwidth,clip, trim=1cm 11cm 7.5cm 7.5cm]{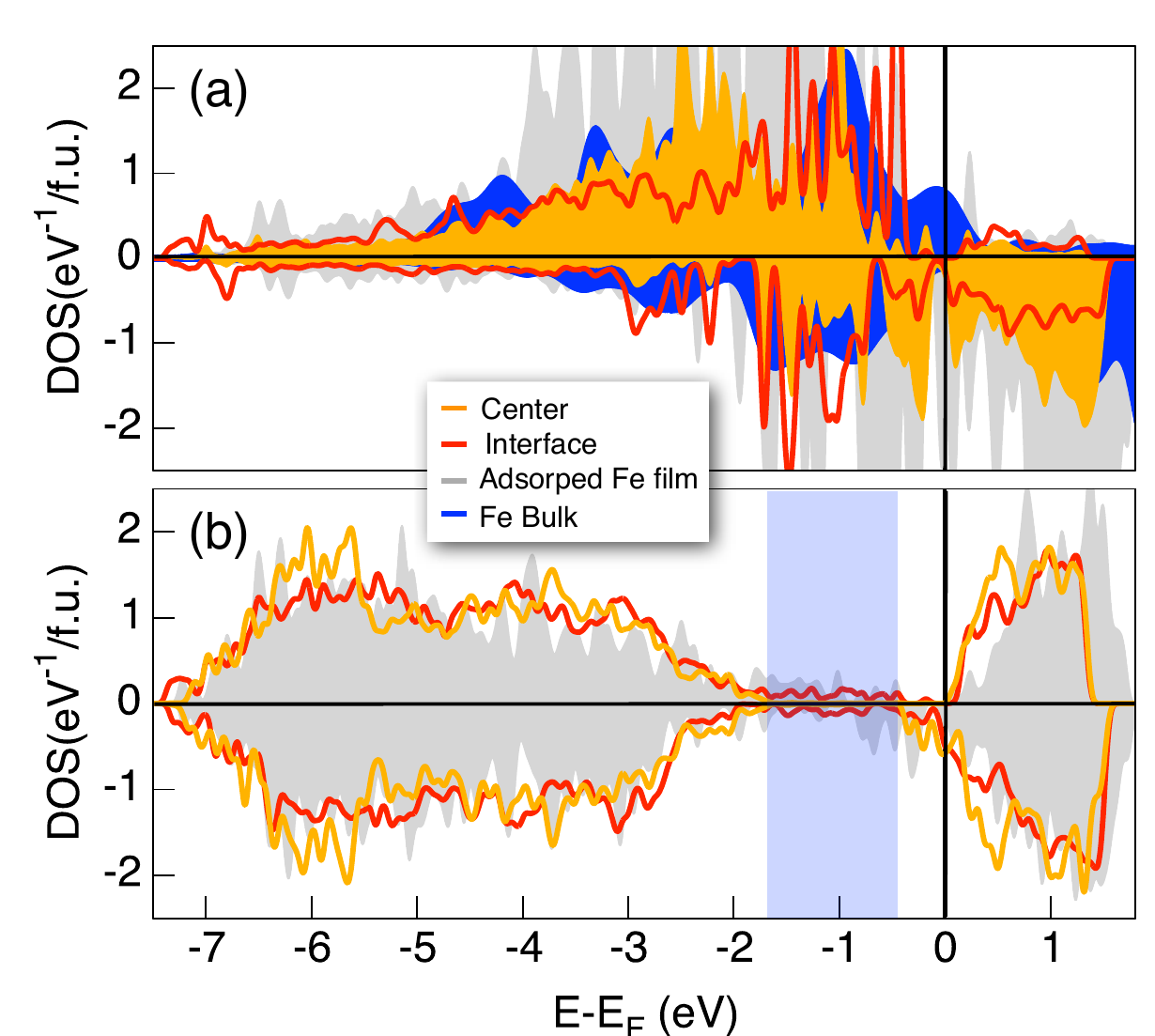}
  \caption{(Color online) Density of states of (a) Fe and (b) TiO$_2$ of the adsorbed Fe$_{14}$ cluster (Fig.~\ref{fig:cluster}(d)). For comparison results for a flat film (Fig.~\ref{fig:layers2}(a)) and bulk Fe are given.}
  \label{fig:doskonv}
 \end{figure}
 mode is likely, if clusters instead of single Fe atoms are deposited onto the surface.\\
The DOS of the adsorbed Fe$_{14}$ cluster agrees qualitatively  with the electronic structure discussed in Sec.~\ref{sec:single}, see Fig.~\ref{fig:doskonv}.
On the one hand, the lowest conduction band states are occupied in the entire film because of the charge spill-out. 
On the other hand, localized MIGs are induced in the shaded energy interval, see  Fig.~\ref{fig:doskonv}(b). In comparison with the adsorption of single atoms, these interface states are broadened to  a "defect" band which fills the gap and the energy intervals of MIGs I and II overlap.\\
Due to the compact cluster which is formed by relaxation, the Fe DOS is more similar to the bulk DOS as in case of flat films and bulk like bands form, especially for the central atom of the Fe cluster. However, no pronounced Fermi edge exists in agreement with experiment.\cite{Nakajima}
\section{Magnetic properties and magneto-electrical coupling of rutile (110)/ iron}
\label{sec:mag}
The adsorption of Fe on the rutile surface incorporates the spin degree of freedom and new functionalities may be thought of. Table~\ref{tab:bader} lists the magnetic moments for single adsorbed atoms.
For all adsorption sites the  magnetic moment is reduced compared to the value of the free atom (4~$\mu_{\text{B}}$).
For positions 1-6, the Fe-TiO$_2$ interaction reduces the Fe moments by $\sim20$\% whereas the weak bonding in case of position 7 results in a reduction of only 10\%. 
Most notably, all Fe moments are larger than the bulk value of of 2.2~$\mu_{\text{B}}$.
If the Fe coverage increases to 0.4~ML, 0.8~ML, 2.4~ML, and 4~ML in a layer-wise growth mode, mean magnetic moments of  3.0, 2.9, 2.8, and 2.7~$\mu_{\text{B}}$ are obtained in the simulations, i.e. the magnetic moments slowly converge towards the bulk value.\footnote{Besides a small enhancement of the moments due to the tensile strain.}\\ 
For adsorbed Fe$_{13}$ and Fe$_{14}$ clusters, the interface atoms without strong interface bonds possess magnetic moments of 2.9~$\mu_{\text B}$ similar to the moments of  3.1 and 2.8 obtained for the free Fe$_{13}$ cluster in its high- and low spin state, respectively.\cite{Bobadova} 
 The moments of the central atoms are reduced with respect to the high-spin phase of the free cluster (2.8~$\mu_{\text B}$ \cite{Bobadova}) and the 
 low-spin state with an antiferromagnetic alignment of the central atom is not stable for adsorbed Fe$_{14}$ clusters. 
If strong Fe-O and Fe-Ti$_{5c}$ bonds form the corresponding magnetic moments are reduced whereas the moments are not modified in case of weak interface bonds, see Fig.~\ref{fig:cluster}. An exception is the ideal icosahedron (g) for which the large oxygen coordination of the vertex atom results in an anti-ferromagnetic alignment of this atom. \\
In summary, the magnetic moment of  Fe is stable in the vicinity of the TiO$_2$ surface which is essential for spintronics applications.\\
Experimentally, a magneto-electrical (ME) coupling has been determined for TiO$_2$-Fe composites.\cite{Yoon}  Such coupling can be mediated by interface 
\begin{figure}
\includegraphics[width=0.5\textwidth, clip, trim=4cm 9cm 9cm 7.5cm]{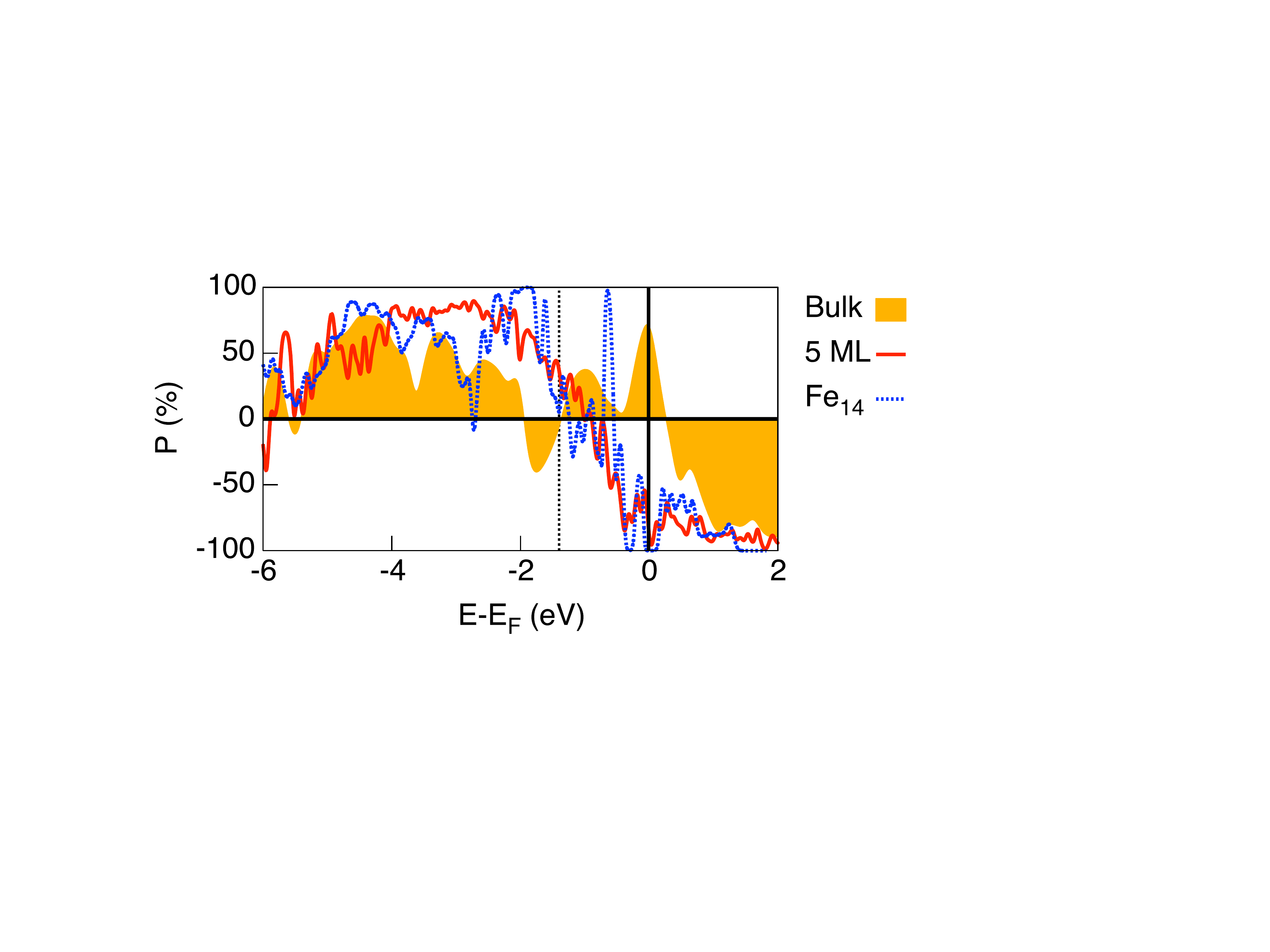}
\caption{Spin-polarization of Fe  in the first Fe adlayer (Fig.~\ref{fig:layers2}(b)) and in a deposited Fe$_{14}$ cluster (Fig.~\ref{fig:cluster}(b)) on rutile (110) in comparison to the spin-polarization of bulk Fe.\label{fig:pol}}
\end{figure}
hybridization\cite{duan} or spin-accumulation.\cite{Rondinelli} 
The discussed pathologic band alignment and the resulting charge spill-out forbid a quantitative estimation of the coupling in the present investigation.\cite{Stengel2}
However, a spin-polarization at the Fermi level of 57\% is obtained in bulk Fe,\cite{intermag} which is most likely sufficient for a spin-accumulation during the screening of the interface charge which would appear under an applied electrical field.
Figure~\ref{fig:pol} illustrates the calculated spin-polarization of the composite material. At  the Fermi level, the polarization is reversed with respect to the bulk material and a large polarization of -90\% is obtained, which would allow for spin accumulation at the interface.\\ 
Complementary to the expected charge transfer and spin-accumulation, a significant ME coupling can be expected because of the magnetic properties of the Fe-TiO$_2$ hybrid states analogously to the discussion of BaTiO$_3$/Fe in Ref.~\onlinecite{duan}. 
Besides the modification of the Fe moments at the interface, magnetic moments are induced in the TiO$_2$ film as both the 
 spurious free charges and the MIGs at the interface are spin-polarized, compare Figs.~\ref{fig:singledos}-\ref{fig:densedos} and  Tab.~\ref{tab:bader}.
 As the lowest Ti$_{5c}$ $d$-states align with the minority Fe states, small magnetic moments are thus induced at the Ti$_{5c}$ atoms in the whole film, which are antiparallel to the Fe moments.\\
For 0.1~ML Fe on adsorption position 1 the magnetic properties of the bound charges can be separated from these spurious charges by integrating the spin-polarized DOS up to $-0.4$~eV, compare Fig.~\ref{fig:singledos2}.
Within this energy interval, only magnetic moments in the direct interface exceed the accuracy of  $0.01~\mu_{\text{B}}$ and the induced moments can be 
traced back to the hybridization with MIGs I and II. Magnetic moments of $\sim0.1$~$\mu_{\text{B}}$ are induced at O$_{\text{b}}$ and O$_{\text{p}}$ atoms for which a Fe-O hybridization with the Ti peak I takes place.
 The Ti$_{6c}$ atom with a Fe-Ti distance of 3.2~{\AA} and the next-nearest Ti$_{6c}$ atoms hybridize with the minority state II  and magnetic moments of $-0.1$~$\mu_{\text{B}}$ and $-$0.05~$\mu_{\text{B}}$ are induced, respectively.\\
Both, the Fe moments at the interface and the induced Ti moments are sensitive to the interface configuration and are thus most likely modified if polar displacements are induced in the film.
For instance,  no hybridization between the Fe state II and  Ti appears for position 2 and thus no magnetic moments are induced at the interface Ti atoms, see Fig.~\ref{fig:singledos2}.\\
For a coverage of 0.4~ML, states are induced within the whole band gap, see Fig.~\ref{fig:densedos}.
However, the separation of the magnetization up to $-0.4$~eV is still a good approximation for position 1, as the surface states in this energy interval decay rapidly with the interface distance. In this energy interval magnetic moments of 0.16,$-0.25$, and 0.07~$\mu_{\text{B}}$ are induced at O$_{\text{b}}$, Ti$_{6c}$, and O$_{\text{p}}$ atoms, respectively.
The induced moments in the TiO$_2$ film are increased as more polarized interface states are occupied, {\it{cf}}.\  the induced moments for 0.1~ML coverage in Tab.~\ref{tab:bader}.\\ 
In summary, magnetic moments are induced by the interface bonds, which are sensitive to the Fe-Ti and Fe-O distances and thus a similar ME-coupling as in BaTiO$_3$/Fe can be expected. For a more quantitative analysis of the ME-coupling strength, accurate hybrid functionals and finite electrical fields must be considered in further calculations.
\section{Conclusions and Outlook}
\label{sec:conclusion}
We have investigated the adsorption of Fe on the rutile (110)-surface by means of {\it{ab inito}} calculations. First, the properties of the free rutile (110)-surface have been reviewed. We could show that the slow convergence of the surface properties with the film thickness can be improved if the bottom layers are fixed to their bulk position. Already for a film thickness of three layers the atomic and electronic structure of the direct surface is mainly converged.\\
Our investigation of the adsorption of single Fe atoms shows that mainly ionic Fe-interface bonds form. The adsorption next to the topmost bridging  O atoms is found most favorable as this atom has an enhanced electronegativity at the free surface. The Fe-O bonds mediate a charge transfer to the neighboring surface Ti atoms which are partly reduced. By the formation of Fe-Ti and Fe-O hybrid states, the band gap is reduced which is favorable for photocathalytic applications.\\
The comparative study of the interface structure for different start configurations and different amount of deposited Fe confirms the wide spread of experimentally found interface structures.
 On the one hand, a large interface coupling and good surface wetting can be obtained if single Fe atoms are deposited. For this setup we found short Fe-interface bonds with a large charge transfer from the Fe atom into the surface.
   On the other hand, Fe-Fe bonds reduce the interface coupling for deposited clusters resulting in a low surface wetting and a three-dimensional interface structure.
 Furthermore the interface structure depends on the adsorption of the first atoms which reflects the influence of the experimental growth process.\\
 Besides the structural and electronic properties, we also discussed the magnetic properties of the interface. We found stable magnetic Fe moments and a large spin polarization of the adsorped Fe atoms. In addition magnetic moments are induced at the Ti and O atoms in the interface. Most notably, the magnetic interface moments depend on the actual interface geometry and thus a hybridization mediated magneto-electrical coupling is very likely.\\
However, more sophisticated approximations for the exchange and correlation potential such as hybrid functionals have to be used in the future in order to obtain the correct electronic gap.
In this case a quantitive estimation of the photocathalytic properties and the magneto-electric coupling will be possible.
\section{Acknowledgments}
This work was supported by the Deutsche Forschungsgemeinschaft (SFB 445, SFB 491, SPP 1599). The authors thank the John von Neumann Institute of Computing (NIC) and the J\"ulich Supercomputing Center (JSC), as well as the Center of Computational Science and Simulation (CCSS) of the University Duisburg-Essen for computational resources.

\begin{thebibliography}{53}%
\makeatletter
\providecommand \@ifxundefined [1]{%
 \@ifx{#1\undefined}
}%
\providecommand \@ifnum [1]{%
 \ifnum #1\expandafter \@firstoftwo
 \else \expandafter \@secondoftwo
 \fi
}%
\providecommand \@ifx [1]{%
 \ifx #1\expandafter \@firstoftwo
 \else \expandafter \@secondoftwo
 \fi
}%
\providecommand \natexlab [1]{#1}%
\providecommand \enquote  [1]{``#1''}%
\providecommand \bibnamefont  [1]{#1}%
\providecommand \bibfnamefont [1]{#1}%
\providecommand \citenamefont [1]{#1}%
\providecommand \href@noop [0]{\@secondoftwo}%
\providecommand \href [0]{\begingroup \@sanitize@url \@href}%
\providecommand \@href[1]{\@@startlink{#1}\@@href}%
\providecommand \@@href[1]{\endgroup#1\@@endlink}%
\providecommand \@sanitize@url [0]{\catcode `\\12\catcode `\$12\catcode
  `\&12\catcode `\#12\catcode `\^12\catcode `\_12\catcode `\%12\relax}%
\providecommand \@@startlink[1]{}%
\providecommand \@@endlink[0]{}%
\providecommand \url  [0]{\begingroup\@sanitize@url \@url }%
\providecommand \@url [1]{\endgroup\@href {#1}{\urlprefix }}%
\providecommand \urlprefix  [0]{URL }%
\providecommand \Eprint [0]{\href }%
\providecommand \doibase [0]{http://dx.doi.org/}%
\providecommand \selectlanguage [0]{\@gobble}%
\providecommand \bibinfo  [0]{\@secondoftwo}%
\providecommand \bibfield  [0]{\@secondoftwo}%
\providecommand \translation [1]{[#1]}%
\providecommand \BibitemOpen [0]{}%
\providecommand \bibitemStop [0]{}%
\providecommand \bibitemNoStop [0]{.\EOS\space}%
\providecommand \EOS [0]{\spacefactor3000\relax}%
\providecommand \BibitemShut  [1]{\csname bibitem#1\endcsname}%
\let\auto@bib@innerbib\@empty
\bibitem [{\citenamefont {Diebold}\ \emph {et~al.}(1995)\citenamefont
  {Diebold}, \citenamefont {Pan},\ and\ \citenamefont {Madey}}]{Diebold}%
  \BibitemOpen
  \bibfield  {author} {\bibinfo {author} {\bibfnamefont {U.}~\bibnamefont
  {Diebold}}, \bibinfo {author} {\bibfnamefont {J.~M.}\ \bibnamefont {Pan}}, \
  and\ \bibinfo {author} {\bibfnamefont {T.~E.}\ \bibnamefont {Madey}},\
  }\href@noop {} {\bibfield  {journal} {\bibinfo  {journal} {Surf. Sci.}\
  }\textbf {\bibinfo {volume} {331}},\ \bibinfo {pages} {845} (\bibinfo {year}
  {1995})}\BibitemShut {NoStop}%
\bibitem [{\citenamefont {Zhang}\ \emph {et~al.}(1992)\citenamefont {Zhang},
  \citenamefont {Hashimoto},\ and\ \citenamefont {Joy}}]{Zhang}%
  \BibitemOpen
  \bibfield  {author} {\bibinfo {author} {\bibfnamefont {X.}~\bibnamefont
  {Zhang}}, \bibinfo {author} {\bibfnamefont {T.}~\bibnamefont {Hashimoto}}, \
  and\ \bibinfo {author} {\bibfnamefont {D.}~\bibnamefont {Joy}},\ }\href@noop
  {} {\bibfield  {journal} {\bibinfo  {journal} {Appl. Phys. Lett.}\ }\textbf
  {\bibinfo {volume} {60}} (\bibinfo {year} {1992})}\BibitemShut {NoStop}%
\bibitem [{\citenamefont {Henderson}(2011)}]{Henderson}%
  \BibitemOpen
  \bibfield  {author} {\bibinfo {author} {\bibfnamefont {M.~A.}\ \bibnamefont
  {Henderson}},\ }\href@noop {} {\bibfield  {journal} {\bibinfo  {journal}
  {Surf. Sci. Rep.}\ }\textbf {\bibinfo {volume} {66}},\ \bibinfo {pages} {185}
  (\bibinfo {year} {2011})}\BibitemShut {NoStop}%
\bibitem [{\citenamefont {Fujishima}\ \emph {et~al.}(2004)\citenamefont
  {Fujishima}, \citenamefont {Zhang},\ and\ \citenamefont {Tryk}}]{Fujishima}%
  \BibitemOpen
  \bibfield  {author} {\bibinfo {author} {\bibfnamefont {A.}~\bibnamefont
  {Fujishima}}, \bibinfo {author} {\bibfnamefont {X.}~\bibnamefont {Zhang}}, \
  and\ \bibinfo {author} {\bibfnamefont {D.~A.}\ \bibnamefont {Tryk}},\
  }\href@noop {} {\bibfield  {journal} {\bibinfo  {journal} {Surf. Sci. Rep.}\
  }\textbf {\bibinfo {volume} {63}},\ \bibinfo {pages} {512 } (\bibinfo {year}
  {2004})}\BibitemShut {NoStop}%
\bibitem [{\citenamefont {Gr\"unebohm}\ \emph {et~al.}(2012)\citenamefont
  {Gr\"unebohm}, \citenamefont {Siewert}, \citenamefont {Entel},\ and\
  \citenamefont {Ederer}}]{Bordeaux}%
  \BibitemOpen
  \bibfield  {author} {\bibinfo {author} {\bibfnamefont {A.}~\bibnamefont
  {Gr\"unebohm}}, \bibinfo {author} {\bibfnamefont {M.}~\bibnamefont
  {Siewert}}, \bibinfo {author} {\bibfnamefont {P.}~\bibnamefont {Entel}}, \
  and\ \bibinfo {author} {\bibfnamefont {C.}~\bibnamefont {Ederer}},\
  }\href@noop {} {\bibfield  {journal} {\bibinfo  {journal} {Ferroelectrics}\
  }\textbf {\bibinfo {volume} {429}},\ \bibinfo {pages} {31} (\bibinfo {year}
  {2012})}\BibitemShut {NoStop}%
\bibitem [{\citenamefont {Teoh}\ \emph {et~al.}(2006)\citenamefont {Teoh},
  \citenamefont {Amal}, \citenamefont {M{\"a}dler},\ and\ \citenamefont
  {Pratsinis}}]{Teoh}%
  \BibitemOpen
  \bibfield  {author} {\bibinfo {author} {\bibfnamefont {W.~Y.}\ \bibnamefont
  {Teoh}}, \bibinfo {author} {\bibfnamefont {R.}~\bibnamefont {Amal}}, \bibinfo
  {author} {\bibfnamefont {L.}~\bibnamefont {M{\"a}dler}}, \ and\ \bibinfo
  {author} {\bibfnamefont {S.~E.}\ \bibnamefont {Pratsinis}},\ }\href@noop {}
  {\bibfield  {journal} {\bibinfo  {journal} {Catal. Today}\ }\textbf {\bibinfo
  {volume} {120}},\ \bibinfo {pages} {203} (\bibinfo {year}
  {2006})}\BibitemShut {NoStop}%
\bibitem [{\citenamefont {Noln}(2011)}]{Nolan}%
  \BibitemOpen
  \bibfield  {author} {\bibinfo {author} {\bibfnamefont {M.}~\bibnamefont
  {Noln}},\ }\href@noop {} {\bibfield  {journal} {\bibinfo  {journal} {Phys.
  Chem. Chem. Phys.}\ }\textbf {\bibinfo {volume} {13}},\ \bibinfo {pages}
  {18194} (\bibinfo {year} {2011})}\BibitemShut {NoStop}%
\bibitem [{\citenamefont {Schierbaum}\ \emph {et~al.}(1993)\citenamefont
  {Schierbaum}, \citenamefont {Wei-Xing}, \citenamefont {Fischer},\ and\
  \citenamefont {G\"obel}}]{Schierbaum}%
  \BibitemOpen
  \bibfield  {author} {\bibinfo {author} {\bibfnamefont {K.~D.}\ \bibnamefont
  {Schierbaum}}, \bibinfo {author} {\bibfnamefont {X.}~\bibnamefont
  {Wei-Xing}}, \bibinfo {author} {\bibfnamefont {S.}~\bibnamefont {Fischer}}, \
  and\ \bibinfo {author} {\bibfnamefont {W.}~\bibnamefont {G\"obel}},\
  }\href@noop {} {\emph {\bibinfo {title} {Adsorption on Ordered Surfaces of
  Ionic Solids and Thin Films}}}\ (\bibinfo  {publisher} {Springer, Berlin},\
  \bibinfo {year} {1993})\BibitemShut {NoStop}%
\bibitem [{\citenamefont {O'Regan}\ and\ \citenamefont
  {Gr\"atzel}(1991)}]{Graetzel}%
  \BibitemOpen
  \bibfield  {author} {\bibinfo {author} {\bibfnamefont {B.}~\bibnamefont
  {O'Regan}}\ and\ \bibinfo {author} {\bibfnamefont {M.}~\bibnamefont
  {Gr\"atzel}},\ }\href@noop {} {\bibfield  {journal} {\bibinfo  {journal}
  {Nature}\ }\textbf {\bibinfo {volume} {353}},\ \bibinfo {pages} {737}
  (\bibinfo {year} {1991})}\BibitemShut {NoStop}%
\bibitem [{\citenamefont {Nobile~Jr.}\ \emph {et~al.}(1991)\citenamefont
  {Nobile~Jr.}, \citenamefont {Van~Brunt},\ and\ \citenamefont
  {Dawis~Jr.}}]{Nobile}%
  \BibitemOpen
  \bibfield  {author} {\bibinfo {author} {\bibfnamefont {A.}~\bibnamefont
  {Nobile~Jr.}}, \bibinfo {author} {\bibfnamefont {V.}~\bibnamefont
  {Van~Brunt}}, \ and\ \bibinfo {author} {\bibfnamefont {M.~W.}\ \bibnamefont
  {Dawis~Jr.}},\ }\href@noop {} {\bibfield  {journal} {\bibinfo  {journal} {J.
  Catal.}\ }\textbf {\bibinfo {volume} {127}},\ \bibinfo {pages} {227}
  (\bibinfo {year} {1991})}\BibitemShut {NoStop}%
\bibitem [{\citenamefont {Matsumoto}\ \emph {et~al.}(2001)\citenamefont
  {Matsumoto}, \citenamefont {Murakami}, \citenamefont {Shono}, \citenamefont
  {Hasegawa}, \citenamefont {Fukumura}, \citenamefont {Kawasaki}, \citenamefont
  {Ahmet}, \citenamefont {Chikyow}, \citenamefont {Koshihara},\ and\
  \citenamefont {Koinuma}}]{Co-DMS}%
  \BibitemOpen
  \bibfield  {author} {\bibinfo {author} {\bibfnamefont {Y.}~\bibnamefont
  {Matsumoto}}, \bibinfo {author} {\bibfnamefont {M.}~\bibnamefont {Murakami}},
  \bibinfo {author} {\bibfnamefont {T.}~\bibnamefont {Shono}}, \bibinfo
  {author} {\bibfnamefont {T.}~\bibnamefont {Hasegawa}}, \bibinfo {author}
  {\bibfnamefont {M.}~\bibnamefont {Fukumura}}, \bibinfo {author}
  {\bibfnamefont {M.}~\bibnamefont {Kawasaki}}, \bibinfo {author}
  {\bibfnamefont {P.}~\bibnamefont {Ahmet}}, \bibinfo {author} {\bibfnamefont
  {T.}~\bibnamefont {Chikyow}}, \bibinfo {author} {\bibfnamefont
  {S.}~\bibnamefont {Koshihara}}, \ and\ \bibinfo {author} {\bibfnamefont
  {H.}~\bibnamefont {Koinuma}},\ }\href@noop {} {\bibfield  {journal} {\bibinfo
   {journal} {Science}\ }\textbf {\bibinfo {volume} {291}},\ \bibinfo {pages}
  {854} (\bibinfo {year} {2001})}\BibitemShut {NoStop}%
\bibitem [{\citenamefont {Ishikawa}\ and\ \citenamefont
  {Uemori}(1999)}]{Ishikawa}%
  \BibitemOpen
  \bibfield  {author} {\bibinfo {author} {\bibfnamefont {K.}~\bibnamefont
  {Ishikawa}}\ and\ \bibinfo {author} {\bibfnamefont {T.}~\bibnamefont
  {Uemori}},\ }\href {\doibase 10.1103/PhysRevB.60.11841} {\bibfield  {journal}
  {\bibinfo  {journal} {Phys. Rev. B}\ }\textbf {\bibinfo {volume} {60}},\
  \bibinfo {pages} {11841} (\bibinfo {year} {1999})}\BibitemShut {NoStop}%
\bibitem [{\citenamefont {Diebold}(2003)}]{Diebold2}%
  \BibitemOpen
  \bibfield  {author} {\bibinfo {author} {\bibfnamefont {U.}~\bibnamefont
  {Diebold}},\ }\href@noop {} {\bibfield  {journal} {\bibinfo  {journal} {Surf.
  Sci. Rep.}\ }\textbf {\bibinfo {volume} {48}},\ \bibinfo {pages} {53}
  (\bibinfo {year} {2003})}\BibitemShut {NoStop}%
\bibitem [{\citenamefont {Hu}\ \emph {et~al.}(2002)\citenamefont {Hu},
  \citenamefont {Noda},\ and\ \citenamefont {Komiyama}}]{Hu}%
  \BibitemOpen
  \bibfield  {author} {\bibinfo {author} {\bibfnamefont {M.}~\bibnamefont
  {Hu}}, \bibinfo {author} {\bibfnamefont {S.}~\bibnamefont {Noda}}, \ and\
  \bibinfo {author} {\bibnamefont {Komiyama}},\ }\href@noop {} {\bibfield
  {journal} {\bibinfo  {journal} {Surf. Sci.}\ }\textbf {\bibinfo {volume}
  {513}},\ \bibinfo {pages} {530} (\bibinfo {year} {2002})}\BibitemShut
  {NoStop}%
\bibitem [{\citenamefont {Pan}\ and\ \citenamefont {Madey}(1993)}]{Pan}%
  \BibitemOpen
  \bibfield  {author} {\bibinfo {author} {\bibfnamefont {J.-M.}\ \bibnamefont
  {Pan}}\ and\ \bibinfo {author} {\bibfnamefont {T.~E.}\ \bibnamefont
  {Madey}},\ }\href {\doibase 10.1116/1.578476} {\bibfield  {journal} {\bibinfo
   {journal} {J. Vac. Sci. Technol. A}\ }\textbf {\bibinfo {volume} {11}},\
  \bibinfo {pages} {1667} (\bibinfo {year} {1993})}\BibitemShut {NoStop}%
\bibitem [{\citenamefont {Nakajima}\ \emph {et~al.}(2004)\citenamefont
  {Nakajima}, \citenamefont {Kato}, \citenamefont {Okazaki},\ and\
  \citenamefont {Sakisaka}}]{Nakajima}%
  \BibitemOpen
  \bibfield  {author} {\bibinfo {author} {\bibfnamefont {N.}~\bibnamefont
  {Nakajima}}, \bibinfo {author} {\bibfnamefont {H.}~\bibnamefont {Kato}},
  \bibinfo {author} {\bibfnamefont {T.}~\bibnamefont {Okazaki}}, \ and\
  \bibinfo {author} {\bibfnamefont {Y.}~\bibnamefont {Sakisaka}},\ }\href@noop
  {} {\bibfield  {journal} {\bibinfo  {journal} {Surf. Sci.}\ }\textbf
  {\bibinfo {volume} {561}},\ \bibinfo {pages} {79} (\bibinfo {year}
  {2004})}\BibitemShut {NoStop}%
\bibitem [{\citenamefont {Asaduzzaman}\ and\ \citenamefont
  {Kr\"uger}(2007)}]{Asaduzzaman2}%
  \BibitemOpen
  \bibfield  {author} {\bibinfo {author} {\bibfnamefont {A.~M.}\ \bibnamefont
  {Asaduzzaman}}\ and\ \bibinfo {author} {\bibfnamefont {P.}~\bibnamefont
  {Kr\"uger}},\ }\href {\doibase 10.1103/PhysRevB.76.115412} {\bibfield
  {journal} {\bibinfo  {journal} {Phys. Rev. B}\ }\textbf {\bibinfo {volume}
  {76}},\ \bibinfo {pages} {115412} (\bibinfo {year} {2007})}\BibitemShut
  {NoStop}%
\bibitem [{\citenamefont {Murugan}\ \emph {et~al.}(2006)\citenamefont
  {Murugan}, \citenamefont {Kumar},\ and\ \citenamefont {Kawazoe}}]{Murugan}%
  \BibitemOpen
  \bibfield  {author} {\bibinfo {author} {\bibfnamefont {P.}~\bibnamefont
  {Murugan}}, \bibinfo {author} {\bibfnamefont {V.}~\bibnamefont {Kumar}}, \
  and\ \bibinfo {author} {\bibfnamefont {Y.}~\bibnamefont {Kawazoe}},\ }\href
  {\doibase 10.1103/PhysRevB.73.075401} {\bibfield  {journal} {\bibinfo
  {journal} {Phys. Rev. B}\ }\textbf {\bibinfo {volume} {73}},\ \bibinfo
  {pages} {075401} (\bibinfo {year} {2006})}\BibitemShut {NoStop}%
\bibitem [{\citenamefont {Giordano}\ \emph {et~al.}(2001)\citenamefont
  {Giordano}, \citenamefont {Paccioni}, \citenamefont {Bredow},\ and\
  \citenamefont {Sanz}}]{Giordano}%
  \BibitemOpen
  \bibfield  {author} {\bibinfo {author} {\bibfnamefont {L.}~\bibnamefont
  {Giordano}}, \bibinfo {author} {\bibfnamefont {G.}~\bibnamefont {Paccioni}},
  \bibinfo {author} {\bibfnamefont {T.}~\bibnamefont {Bredow}}, \ and\ \bibinfo
  {author} {\bibfnamefont {J.~F.}\ \bibnamefont {Sanz}},\ }\href@noop {}
  {\bibfield  {journal} {\bibinfo  {journal} {Surf. Sci.}\ }\textbf {\bibinfo
  {volume} {471}},\ \bibinfo {pages} {21} (\bibinfo {year} {2001})}\BibitemShut
  {NoStop}%
\bibitem [{\citenamefont {Asaduzzaman}\ and\ \citenamefont
  {Kr\"uger}(2008)}]{Asaduzzaman}%
  \BibitemOpen
  \bibfield  {author} {\bibinfo {author} {\bibfnamefont {A.}~\bibnamefont
  {Asaduzzaman}}\ and\ \bibinfo {author} {\bibfnamefont {P.}~\bibnamefont
  {Kr\"uger}},\ }\href@noop {} {\bibfield  {journal} {\bibinfo  {journal} {J.
  Phys. Chem. C}\ }\textbf {\bibinfo {volume} {112}},\ \bibinfo {pages} {19616}
  (\bibinfo {year} {2008})}\BibitemShut {NoStop}%
\bibitem [{\citenamefont {Duan}\ \emph {et~al.}(2006)\citenamefont {Duan},
  \citenamefont {Jaswal},\ and\ \citenamefont {Tsymbal}}]{duan}%
  \BibitemOpen
  \bibfield  {author} {\bibinfo {author} {\bibfnamefont {C.-G.}\ \bibnamefont
  {Duan}}, \bibinfo {author} {\bibfnamefont {S.~S.}\ \bibnamefont {Jaswal}}, \
  and\ \bibinfo {author} {\bibfnamefont {E.~Y.}\ \bibnamefont {Tsymbal}},\
  }\href {\doibase 10.1103/PhysRevLett.97.047201} {\bibfield  {journal}
  {\bibinfo  {journal} {Phys. Rev. Lett.}\ }\textbf {\bibinfo {volume} {97}},\
  \bibinfo {eid} {047201} (\bibinfo {year} {2006})}\BibitemShut {NoStop}%
\bibitem [{\citenamefont {Kresse}\ and\ \citenamefont
  {Furthm\"uller}(1996)}]{Kresse1}%
  \BibitemOpen
  \bibfield  {author} {\bibinfo {author} {\bibfnamefont {G.}~\bibnamefont
  {Kresse}}\ and\ \bibinfo {author} {\bibfnamefont {J.}~\bibnamefont
  {Furthm\"uller}},\ }\href@noop {} {\bibfield  {journal} {\bibinfo  {journal}
  {Phys. Rev. B}\ }\textbf {\bibinfo {volume} {54}},\ \bibinfo {pages} {11169}
  (\bibinfo {year} {1996})}\BibitemShut {NoStop}%
\bibitem [{\citenamefont {Bl\"ochl}(1994)}]{Blochl}%
  \BibitemOpen
  \bibfield  {author} {\bibinfo {author} {\bibfnamefont {P.~E.}\ \bibnamefont
  {Bl\"ochl}},\ }\href@noop {} {\bibfield  {journal} {\bibinfo  {journal}
  {Phys. Rev. B}\ }\textbf {\bibinfo {volume} {50}},\ \bibinfo {pages} {17953}
  (\bibinfo {year} {1994})}\BibitemShut {NoStop}%
\bibitem [{\citenamefont {Perdew}\ \emph {et~al.}(1996)\citenamefont {Perdew},
  \citenamefont {Burke},\ and\ \citenamefont {Ernzerhof}}]{PBE}%
  \BibitemOpen
  \bibfield  {author} {\bibinfo {author} {\bibfnamefont {J.~P.}\ \bibnamefont
  {Perdew}}, \bibinfo {author} {\bibfnamefont {K.}~\bibnamefont {Burke}}, \
  and\ \bibinfo {author} {\bibfnamefont {M.}~\bibnamefont {Ernzerhof}},\
  }\href@noop {} {\bibfield  {journal} {\bibinfo  {journal} {Phys. Rev. Lett.}\
  }\textbf {\bibinfo {volume} {77}},\ \bibinfo {pages} {3865} (\bibinfo {year}
  {1996})}\BibitemShut {NoStop}%
\bibitem [{\citenamefont {Calzado}\ \emph {et~al.}(2008)\citenamefont
  {Calzado}, \citenamefont {Hern\'andez},\ and\ \citenamefont
  {Sanz}}]{Calzado}%
  \BibitemOpen
  \bibfield  {author} {\bibinfo {author} {\bibfnamefont {C.~J.}\ \bibnamefont
  {Calzado}}, \bibinfo {author} {\bibfnamefont {N.~C.}\ \bibnamefont
  {Hern\'andez}}, \ and\ \bibinfo {author} {\bibfnamefont {J.~F.}\ \bibnamefont
  {Sanz}},\ }\href {\doibase 10.1103/PhysRevB.77.045118} {\bibfield  {journal}
  {\bibinfo  {journal} {Phys. Rev. B}\ }\textbf {\bibinfo {volume} {77}},\
  \bibinfo {pages} {045118} (\bibinfo {year} {2008})}\BibitemShut {NoStop}%
\bibitem [{\citenamefont {Dudarev}\ \emph {et~al.}(1998)\citenamefont
  {Dudarev}, \citenamefont {Botton}, \citenamefont {Savrasov}, \citenamefont
  {Humphreys},\ and\ \citenamefont {Sutton}}]{Dudarev}%
  \BibitemOpen
  \bibfield  {author} {\bibinfo {author} {\bibfnamefont {S.~L.}\ \bibnamefont
  {Dudarev}}, \bibinfo {author} {\bibfnamefont {G.~A.}\ \bibnamefont {Botton}},
  \bibinfo {author} {\bibfnamefont {S.~Y.}\ \bibnamefont {Savrasov}}, \bibinfo
  {author} {\bibfnamefont {C.~J.}\ \bibnamefont {Humphreys}}, \ and\ \bibinfo
  {author} {\bibfnamefont {A.~P.}\ \bibnamefont {Sutton}},\ }\href@noop {}
  {\bibfield  {journal} {\bibinfo  {journal} {Phys. Rev. B}\ }\textbf {\bibinfo
  {volume} {57}},\ \bibinfo {pages} {1505} (\bibinfo {year}
  {1998})}\BibitemShut {NoStop}%
\bibitem [{\citenamefont {Monkhorst}\ and\ \citenamefont
  {Pack}(1976)}]{Monkhorst}%
  \BibitemOpen
  \bibfield  {author} {\bibinfo {author} {\bibfnamefont {H.~J.}\ \bibnamefont
  {Monkhorst}}\ and\ \bibinfo {author} {\bibfnamefont {J.~D.}\ \bibnamefont
  {Pack}},\ }\href@noop {} {\bibfield  {journal} {\bibinfo  {journal} {Phys.
  Rev. B}\ }\textbf {\bibinfo {volume} {13}},\ \bibinfo {pages} {5188}
  (\bibinfo {year} {1976})}\BibitemShut {NoStop}%
\bibitem [{\citenamefont {Muscat}\ \emph {et~al.}(2002)\citenamefont {Muscat},
  \citenamefont {Swamy},\ and\ \citenamefont {Harrison}}]{Muscat}%
  \BibitemOpen
  \bibfield  {author} {\bibinfo {author} {\bibfnamefont {J.}~\bibnamefont
  {Muscat}}, \bibinfo {author} {\bibfnamefont {V.}~\bibnamefont {Swamy}}, \
  and\ \bibinfo {author} {\bibfnamefont {N.~M.}\ \bibnamefont {Harrison}},\
  }\href {\doibase 10.1103/PhysRevB.65.224112} {\bibfield  {journal} {\bibinfo
  {journal} {Phys. Rev. B}\ }\textbf {\bibinfo {volume} {65}},\ \bibinfo
  {pages} {224112} (\bibinfo {year} {2002})}\BibitemShut {NoStop}%
\bibitem [{\citenamefont {Gr\"unebohm}\ \emph {et~al.}(2013)\citenamefont
  {Gr\"unebohm}, \citenamefont {Ederer},\ and\ \citenamefont {Entel}}]{felich}%
  \BibitemOpen
  \bibfield  {author} {\bibinfo {author} {\bibfnamefont {A.}~\bibnamefont
  {Gr\"unebohm}}, \bibinfo {author} {\bibfnamefont {C.}~\bibnamefont {Ederer}},
  \ and\ \bibinfo {author} {\bibfnamefont {P.}~\bibnamefont {Entel}},\
  }\href@noop {} {\bibfield  {journal} {\bibinfo  {journal} {Phys. Rev. B}\
  }\textbf {\bibinfo {volume} {87}},\ \bibinfo {pages} {054110} (\bibinfo
  {year} {2013})}\BibitemShut {NoStop}%
\bibitem [{\citenamefont {Labat}\ \emph {et~al.}(2007)\citenamefont {Labat},
  \citenamefont {Baranek}, \citenamefont {Domain}, \citenamefont {Minot},\ and\
  \citenamefont {Adamo}}]{Labat}%
  \BibitemOpen
  \bibfield  {author} {\bibinfo {author} {\bibfnamefont {F.}~\bibnamefont
  {Labat}}, \bibinfo {author} {\bibfnamefont {P.}~\bibnamefont {Baranek}},
  \bibinfo {author} {\bibfnamefont {C.}~\bibnamefont {Domain}}, \bibinfo
  {author} {\bibfnamefont {C.}~\bibnamefont {Minot}}, \ and\ \bibinfo {author}
  {\bibfnamefont {C.}~\bibnamefont {Adamo}},\ }\href@noop {} {\bibfield
  {journal} {\bibinfo  {journal} {J. Chem. Phys}\ }\textbf {\bibinfo {volume}
  {126}},\ \bibinfo {pages} {154703} (\bibinfo {year} {2007})}\BibitemShut
  {NoStop}%
\bibitem [{\citenamefont {Samara}\ and\ \citenamefont {Peercy}(1973)}]{Samara}%
  \BibitemOpen
  \bibfield  {author} {\bibinfo {author} {\bibfnamefont {G.~A.}\ \bibnamefont
  {Samara}}\ and\ \bibinfo {author} {\bibfnamefont {P.~S.}\ \bibnamefont
  {Peercy}},\ }\href {\doibase 10.1103/PhysRevB.7.1131} {\bibfield  {journal}
  {\bibinfo  {journal} {Phys. Rev. B}\ }\textbf {\bibinfo {volume} {7}},\
  \bibinfo {pages} {1131} (\bibinfo {year} {1973})}\BibitemShut {NoStop}%
\bibitem [{\citenamefont {Traylor}\ \emph {et~al.}(1971)\citenamefont
  {Traylor}, \citenamefont {Smith}, \citenamefont {Nicklow},\ and\
  \citenamefont {Wilkinson}}]{Traylor}%
  \BibitemOpen
  \bibfield  {author} {\bibinfo {author} {\bibfnamefont {J.~G.}\ \bibnamefont
  {Traylor}}, \bibinfo {author} {\bibfnamefont {H.~G.}\ \bibnamefont {Smith}},
  \bibinfo {author} {\bibfnamefont {R.~M.}\ \bibnamefont {Nicklow}}, \ and\
  \bibinfo {author} {\bibfnamefont {M.~K.}\ \bibnamefont {Wilkinson}},\ }\href
  {\doibase 10.1103/PhysRevB.3.3457} {\bibfield  {journal} {\bibinfo  {journal}
  {Phys. Rev. B}\ }\textbf {\bibinfo {volume} {3}},\ \bibinfo {pages} {3457}
  (\bibinfo {year} {1971})}\BibitemShut {NoStop}%
\bibitem [{\citenamefont {Pascual}\ \emph {et~al.}(1978)\citenamefont
  {Pascual}, \citenamefont {Camassel},\ and\ \citenamefont
  {Mathieu}}]{Pascual}%
  \BibitemOpen
  \bibfield  {author} {\bibinfo {author} {\bibfnamefont {J.}~\bibnamefont
  {Pascual}}, \bibinfo {author} {\bibfnamefont {J.}~\bibnamefont {Camassel}}, \
  and\ \bibinfo {author} {\bibfnamefont {H.}~\bibnamefont {Mathieu}},\ }\href
  {\doibase 10.1103/PhysRevB.18.5606} {\bibfield  {journal} {\bibinfo
  {journal} {Phys. Rev. B}\ }\textbf {\bibinfo {volume} {18}},\ \bibinfo
  {pages} {5606} (\bibinfo {year} {1978})}\BibitemShut {NoStop}%
\bibitem [{\citenamefont {Reinhardt}\ and\ \citenamefont
  {He{\ss}}(1994)}]{Reinhardt}%
  \BibitemOpen
  \bibfield  {author} {\bibinfo {author} {\bibfnamefont {P.}~\bibnamefont
  {Reinhardt}}\ and\ \bibinfo {author} {\bibfnamefont {B.~A.}\ \bibnamefont
  {He{\ss}}},\ }\href@noop {} {\bibfield  {journal} {\bibinfo  {journal} {Phys.
  Rev. B}\ }\textbf {\bibinfo {volume} {50}},\ \bibinfo {pages} {12015}
  (\bibinfo {year} {1994})}\BibitemShut {NoStop}%
\bibitem [{\citenamefont {Henkelmann}\ \emph {et~al.}(2006)\citenamefont
  {Henkelmann}, \citenamefont {Arnaldsson},\ and\ \citenamefont
  {J{$\acute{\rm{o}}$}nsson}}]{bader_imp}%
  \BibitemOpen
  \bibfield  {author} {\bibinfo {author} {\bibfnamefont {G.}~\bibnamefont
  {Henkelmann}}, \bibinfo {author} {\bibfnamefont {A.}~\bibnamefont
  {Arnaldsson}}, \ and\ \bibinfo {author} {\bibfnamefont {H.}~\bibnamefont
  {J{$\acute{\rm{o}}$}nsson}},\ }\href@noop {} {\bibfield  {journal} {\bibinfo
  {journal} {Comput. Mater. Sci.}\ }\textbf {\bibinfo {volume} {36}},\ \bibinfo
  {pages} {354} (\bibinfo {year} {2006})}\BibitemShut {NoStop}%
\bibitem [{\citenamefont {Heise}\ \emph {et~al.}(1992)\citenamefont {Heise},
  \citenamefont {Courths},\ and\ \citenamefont {Witzel}}]{Courths}%
  \BibitemOpen
  \bibfield  {author} {\bibinfo {author} {\bibfnamefont {R.}~\bibnamefont
  {Heise}}, \bibinfo {author} {\bibfnamefont {R.}~\bibnamefont {Courths}}, \
  and\ \bibinfo {author} {\bibfnamefont {S.}~\bibnamefont {Witzel}},\
  }\href@noop {} {\bibfield  {journal} {\bibinfo  {journal} {Solid State
  Commun.}\ }\textbf {\bibinfo {volume} {84}},\ \bibinfo {pages} {599}
  (\bibinfo {year} {1992})}\BibitemShut {NoStop}%
\bibitem [{\citenamefont {Bredow}\ \emph {et~al.}(2004)\citenamefont {Bredow},
  \citenamefont {Giordano}, \citenamefont {Cinquini},\ and\ \citenamefont
  {Pacchioni}}]{Bredow}%
  \BibitemOpen
  \bibfield  {author} {\bibinfo {author} {\bibfnamefont {T.}~\bibnamefont
  {Bredow}}, \bibinfo {author} {\bibfnamefont {L.}~\bibnamefont {Giordano}},
  \bibinfo {author} {\bibfnamefont {F.}~\bibnamefont {Cinquini}}, \ and\
  \bibinfo {author} {\bibfnamefont {G.}~\bibnamefont {Pacchioni}},\ }\href
  {\doibase 10.1103/PhysRevB.70.035419} {\bibfield  {journal} {\bibinfo
  {journal} {Phys. Rev. B}\ }\textbf {\bibinfo {volume} {70}},\ \bibinfo
  {pages} {035419} (\bibinfo {year} {2004})}\BibitemShut {NoStop}%
\bibitem [{\citenamefont {Bates}\ \emph {et~al.}(1997)\citenamefont {Bates},
  \citenamefont {Kresse},\ and\ \citenamefont {Gillan}}]{Bates}%
  \BibitemOpen
  \bibfield  {author} {\bibinfo {author} {\bibfnamefont {S.~P.}\ \bibnamefont
  {Bates}}, \bibinfo {author} {\bibfnamefont {G.}~\bibnamefont {Kresse}}, \
  and\ \bibinfo {author} {\bibfnamefont {M.~J.}\ \bibnamefont {Gillan}},\
  }\href {\doibase DOI: 10.1016/S0039-6028(97)00265-3} {\bibfield  {journal}
  {\bibinfo  {journal} {Surf. Sci.}\ }\textbf {\bibinfo {volume} {385}},\
  \bibinfo {pages} {386} (\bibinfo {year} {1997})}\BibitemShut {NoStop}%
\bibitem [{\citenamefont {Schelling}\ \emph {et~al.}(1998)\citenamefont
  {Schelling}, \citenamefont {Yu},\ and\ \citenamefont {Halley}}]{Schelling}%
  \BibitemOpen
  \bibfield  {author} {\bibinfo {author} {\bibfnamefont {P.~K.}\ \bibnamefont
  {Schelling}}, \bibinfo {author} {\bibfnamefont {N.}~\bibnamefont {Yu}}, \
  and\ \bibinfo {author} {\bibfnamefont {J.~W.}\ \bibnamefont {Halley}},\
  }\href@noop {} {\bibfield  {journal} {\bibinfo  {journal} {Phys. Rev. B}\
  }\textbf {\bibinfo {volume} {58}},\ \bibinfo {pages} {1279} (\bibinfo {year}
  {1998})}\BibitemShut {NoStop}%
\bibitem [{\citenamefont {Diebold}\ \emph {et~al.}(1994)\citenamefont
  {Diebold}, \citenamefont {Tao}, \citenamefont {Shinn},\ and\ \citenamefont
  {Madey}}]{Diebold3}%
  \BibitemOpen
  \bibfield  {author} {\bibinfo {author} {\bibfnamefont {U.}~\bibnamefont
  {Diebold}}, \bibinfo {author} {\bibfnamefont {H.-S.}\ \bibnamefont {Tao}},
  \bibinfo {author} {\bibfnamefont {N.~D.}\ \bibnamefont {Shinn}}, \ and\
  \bibinfo {author} {\bibfnamefont {T.~E.}\ \bibnamefont {Madey}},\ }\href
  {\doibase 10.1103/PhysRevB.50.14474} {\bibfield  {journal} {\bibinfo
  {journal} {Phys. Rev. B}\ }\textbf {\bibinfo {volume} {50}},\ \bibinfo
  {pages} {14474} (\bibinfo {year} {1994})}\BibitemShut {NoStop}%
\bibitem [{\citenamefont {Stengel}\ \emph {et~al.}(2011)\citenamefont
  {Stengel}, \citenamefont {Aguado-Puente}, \citenamefont {Spaldin},\ and\
  \citenamefont {Junquera}}]{Stengel2}%
  \BibitemOpen
  \bibfield  {author} {\bibinfo {author} {\bibfnamefont {M.}~\bibnamefont
  {Stengel}}, \bibinfo {author} {\bibfnamefont {P.}~\bibnamefont
  {Aguado-Puente}}, \bibinfo {author} {\bibfnamefont {N.~A.}\ \bibnamefont
  {Spaldin}}, \ and\ \bibinfo {author} {\bibfnamefont {J.}~\bibnamefont
  {Junquera}},\ }\href {\doibase 10.1103/PhysRevB.83.235112} {\bibfield
  {journal} {\bibinfo  {journal} {Phys. Rev. B}\ }\textbf {\bibinfo {volume}
  {83}},\ \bibinfo {pages} {235112} (\bibinfo {year} {2011})}\BibitemShut
  {NoStop}%
\bibitem [{Note1()}]{Note1}%
  \BibitemOpen
  \bibinfo {note} {For some configurations no such unique separation is
  possible, e.g., for position 3, see Fig.~\ref {fig:singledos2}.}\BibitemShut
  {Stop}%
\bibitem [{\citenamefont {Zhou}\ \emph {et~al.}(2008)\citenamefont {Zhou},
  \citenamefont {Talut}, \citenamefont {Potzger}, \citenamefont {Shalimov},
  \citenamefont {Grenzer}, \citenamefont {Skorupa}, \citenamefont {Helm},
  \citenamefont {Fassbender}, \citenamefont {\v{C}i\v{z}m\'{a}r},\ and\
  \citenamefont {Zvyagin}}]{Zhou}%
  \BibitemOpen
  \bibfield  {author} {\bibinfo {author} {\bibfnamefont {S.}~\bibnamefont
  {Zhou}}, \bibinfo {author} {\bibfnamefont {G.}~\bibnamefont {Talut}},
  \bibinfo {author} {\bibfnamefont {K.}~\bibnamefont {Potzger}}, \bibinfo
  {author} {\bibfnamefont {A.}~\bibnamefont {Shalimov}}, \bibinfo {author}
  {\bibfnamefont {J.}~\bibnamefont {Grenzer}}, \bibinfo {author} {\bibfnamefont
  {W.}~\bibnamefont {Skorupa}}, \bibinfo {author} {\bibfnamefont
  {M.}~\bibnamefont {Helm}}, \bibinfo {author} {\bibfnamefont {J.}~\bibnamefont
  {Fassbender}}, \bibinfo {author} {\bibfnamefont {E.}~\bibnamefont
  {\v{C}i\v{z}m\'{a}r}}, \ and\ \bibinfo {author} {\bibfnamefont {S.~A.}\
  \bibnamefont {Zvyagin}},\ }\href@noop {} {\bibfield  {journal} {\bibinfo
  {journal} {J. Appl. Phys.}\ }\textbf {\bibinfo {volume} {103}},\ \bibinfo
  {pages} {083907} (\bibinfo {year} {2008})}\BibitemShut {NoStop}%
\bibitem [{Note2()}]{Note2}%
  \BibitemOpen
  \bibinfo {note} {The Fe positions and the upper TiO$_2$ layers have been
  fully relaxed for different adsorption sites.}\BibitemShut {Stop}%
\bibitem [{\citenamefont {Rollmann}(2007)}]{dissgeorg}%
  \BibitemOpen
  \bibfield  {author} {\bibinfo {author} {\bibfnamefont {G.}~\bibnamefont
  {Rollmann}},\ }\emph {\bibinfo {title} {Ab initio Simulation eisenhaltiger
  Systeme: Vom Festk\"orper zum Cluster}},\ \href@noop {} {Ph.D. thesis},\
  \bibinfo  {school} {Universi\"at Duisburg-Essen} (\bibinfo {year}
  {2007})\BibitemShut {NoStop}%
\bibitem [{Note3()}]{Note3}%
  \BibitemOpen
  \bibinfo {note} {Due to the different symmetries of rutile and Fe
  inequivalent adsorption positions are occupied during the formation of a
  closed film anyway.}\BibitemShut {Stop}%
\bibitem [{\citenamefont {Rollmann}\ \emph {et~al.}(2007)\citenamefont
  {Rollmann}, \citenamefont {Gruner}, \citenamefont {Hucht}, \citenamefont
  {Meyer}, \citenamefont {Entel}, \citenamefont {Tiago},\ and\ \citenamefont
  {Chelikowsky}}]{prlgeorg}%
  \BibitemOpen
  \bibfield  {author} {\bibinfo {author} {\bibfnamefont {G.}~\bibnamefont
  {Rollmann}}, \bibinfo {author} {\bibfnamefont {M.~E.}\ \bibnamefont
  {Gruner}}, \bibinfo {author} {\bibfnamefont {A.}~\bibnamefont {Hucht}},
  \bibinfo {author} {\bibfnamefont {R.}~\bibnamefont {Meyer}}, \bibinfo
  {author} {\bibfnamefont {P.}~\bibnamefont {Entel}}, \bibinfo {author}
  {\bibfnamefont {M.~L.}\ \bibnamefont {Tiago}}, \ and\ \bibinfo {author}
  {\bibfnamefont {J.~R.}\ \bibnamefont {Chelikowsky}},\ }\href {\doibase
  10.1103/PhysRevLett.99.083402} {\bibfield  {journal} {\bibinfo  {journal}
  {Phys. Rev. Lett.}\ }\textbf {\bibinfo {volume} {99}},\ \bibinfo {pages}
  {083402} (\bibinfo {year} {2007})}\BibitemShut {NoStop}%
\bibitem [{Note4()}]{Note4}%
  \BibitemOpen
  \bibinfo {note} {The relaxation has only been performed up to 0.001
  eV/step.}\BibitemShut {Stop}%
\bibitem [{Note5()}]{Note5}%
  \BibitemOpen
  \bibinfo {note} {Besides a small enhancement of the moments due to the
  tensile strain.}\BibitemShut {Stop}%
\bibitem [{\citenamefont {Bobadova-Parvanova}\ \emph
  {et~al.}(2002)\citenamefont {Bobadova-Parvanova}, \citenamefont {Jackson},
  \citenamefont {Srinivas},\ and\ \citenamefont {Horoi}}]{Bobadova}%
  \BibitemOpen
  \bibfield  {author} {\bibinfo {author} {\bibfnamefont {P.}~\bibnamefont
  {Bobadova-Parvanova}}, \bibinfo {author} {\bibfnamefont {K.~A.}\ \bibnamefont
  {Jackson}}, \bibinfo {author} {\bibfnamefont {S.}~\bibnamefont {Srinivas}}, \
  and\ \bibinfo {author} {\bibfnamefont {M.}~\bibnamefont {Horoi}},\ }\href
  {\doibase 10.1103/PhysRevB.66.195402} {\bibfield  {journal} {\bibinfo
  {journal} {Phys. Rev. B}\ }\textbf {\bibinfo {volume} {66}},\ \bibinfo
  {pages} {195402} (\bibinfo {year} {2002})}\BibitemShut {NoStop}%
\bibitem [{\citenamefont {Yoon}\ \emph {et~al.}(2008)\citenamefont {Yoon},
  \citenamefont {Vittoria}, \citenamefont {Srivastava}, \citenamefont {Widom},\
  and\ \citenamefont {Harris}}]{Yoon}%
  \BibitemOpen
  \bibfield  {author} {\bibinfo {author} {\bibfnamefont {S.~D.}\ \bibnamefont
  {Yoon}}, \bibinfo {author} {\bibfnamefont {C.}~\bibnamefont {Vittoria}},
  \bibinfo {author} {\bibfnamefont {Y.}~\bibnamefont {Srivastava}}, \bibinfo
  {author} {\bibfnamefont {A.}~\bibnamefont {Widom}}, \ and\ \bibinfo {author}
  {\bibfnamefont {V.~G.}\ \bibnamefont {Harris}},\ }\href@noop {} {\bibfield
  {journal} {\bibinfo  {journal} {Appl. Phys. Lett.}\ }\textbf {\bibinfo
  {volume} {92}},\ \bibinfo {pages} {042508} (\bibinfo {year}
  {2008})}\BibitemShut {NoStop}%
\bibitem [{\citenamefont {Rondinelli}\ \emph {et~al.}(2008)\citenamefont
  {Rondinelli}, \citenamefont {Stengel},\ and\ \citenamefont
  {Spaldin}}]{Rondinelli}%
  \BibitemOpen
  \bibfield  {author} {\bibinfo {author} {\bibfnamefont {J.~M.}\ \bibnamefont
  {Rondinelli}}, \bibinfo {author} {\bibfnamefont {M.}~\bibnamefont {Stengel}},
  \ and\ \bibinfo {author} {\bibfnamefont {N.~A.}\ \bibnamefont {Spaldin}},\
  }\href@noop {} {\bibfield  {journal} {\bibinfo  {journal} {Nature Nanotech.}\
  }\textbf {\bibinfo {volume} {3}},\ \bibinfo {pages} {46} (\bibinfo {year}
  {2008})}\BibitemShut {NoStop}%
\bibitem [{\citenamefont {Gr{\"u}nebohm}\ \emph {et~al.}(2009)\citenamefont
  {Gr{\"u}nebohm}, \citenamefont {Herper},\ and\ \citenamefont
  {Entel}}]{intermag}%
  \BibitemOpen
  \bibfield  {author} {\bibinfo {author} {\bibfnamefont {A.}~\bibnamefont
  {Gr{\"u}nebohm}}, \bibinfo {author} {\bibfnamefont {H.~C.}\ \bibnamefont
  {Herper}}, \ and\ \bibinfo {author} {\bibfnamefont {P.}~\bibnamefont
  {Entel}},\ }\href@noop {} {\bibfield  {journal} {\bibinfo  {journal} {IEEE
  Trans. Magn.}\ }\textbf {\bibinfo {volume} {45}},\ \bibinfo {pages} {3965}
  (\bibinfo {year} {2009})}\BibitemShut {NoStop}%
\end{thebibliography}
%
\end{document}